\documentclass[aps,prb,twocolumn,showpacs,superscriptaddress,floatfix]{revtex4}
\usepackage{amssymb,amsmath}
\usepackage{graphicx}
\usepackage{multirow}
\usepackage{hyperref}
\usepackage{appendix}
\usepackage{mathtools}
\usepackage{todonotes}
\usepackage{amsmath}
\usepackage{xcolor}
\usepackage{siunitx}

\usepackage{xr}

\graphicspath{{./IMG/}}

\newcommand{\fro}{{Fr\"ohlich~}}

\begin{document}

\title{
\fro polaron effective mass and localization length in cubic
materials: degenerate and anisotropic electronic bands}

\author{Bogdan Guster}
\affiliation{European Theoretical Spectroscopy Facility, Institute of Condensed Matter and Nanosciences, Universit\'{e} catholique de Louvain, Chemin des \'{e}toiles 8, bte L07.03.01, B-1348 Louvain-la-Neuve, Belgium}
\author{Pedro Melo}
\affiliation{NanoMat/Q-Mat/CESAM and European Theoretical Spectroscopy Facility, Universit\'e de Li\`ege (B5), B-4000 Li\`ege, Belgium}
\author{Bradley A. A. Martin}
\affiliation{Department of Physics, Imperial College London, Exhibition Road, London  SW7 2AZ, UK}
\author{V\'eronique Brousseau-Couture}
\affiliation{
D\'epartement de Physique et Regroupement Qu\'eb\'ecois sur les Mat\'eriaux de Pointe, Universit\'e de Montreal, C.P. 6128, Succursale Centre-Ville, Montreal, Canada H3C 3J7
}
\author{Anna Miglio}
\affiliation{European Theoretical Spectroscopy Facility, Institute of Condensed Matter and Nanosciences, Universit\'{e} catholique de Louvain, Chemin des \'{e}toiles 8, bte L07.03.01, B-1348 Louvain-la-Neuve, Belgium}
\author{Matteo Giantomassi}
\affiliation{European Theoretical Spectroscopy Facility, Institute of Condensed Matter and Nanosciences, Universit\'{e} catholique de Louvain, Chemin des \'{e}toiles 8, bte L07.03.01, B-1348 Louvain-la-Neuve, Belgium}
\author{Michel C\^ot\'e}
\affiliation{
D\'epartement de Physique et Regroupement Qu\'eb\'ecois sur les Mat\'eriaux de Pointe, Universit\'e de Montreal, C.P. 6128, Succursale Centre-Ville, Montreal, Canada H3C 3J7
}
\author{Jarvist M. Frost}
\affiliation{Department of Physics, Imperial College London, Exhibition Road, London  SW7 2AZ, UK}
\author{Matthieu J. Verstraete}
\affiliation{NanoMat/Q-Mat/CESAM and European Theoretical Spectroscopy Facility, Universit\'e de Li\`ege (B5), B-4000 Li\`ege, Belgium}
\author{Xavier Gonze}
\affiliation{European Theoretical Spectroscopy Facility, Institute of Condensed Matter and Nanosciences, Universit\'{e} catholique de Louvain, Chemin des \'{e}toiles 8, bte L07.03.01, B-1348 Louvain-la-Neuve, Belgium}\affiliation{Skolkovo Institute of
Science and Technology, Skolkovo Innovation Center, Nobel St. 3, Moscow, 143026, Russia.}

\date{\today}

\begin{abstract}
Polarons, that is, charge carriers correlated with lattice deformations, are ubiquitous quasiparticles in semiconductors, and play an important role in electrical conductivity. To date most theoretical studies of so-called large polarons, in which the lattice can be considered as a continuum, have focused on the original 
\fro model: a simple (non-degenerate) parabolic isotropic electronic band coupled to one dispersionless longitudinal optical phonon branch.
The \fro model allows one to understand characteristics such as polaron formation energy, radius, effective mass and mobility.
Real cubic materials, instead, have electronic band extrema that are often degenerate (e.g. 3-fold degeneracy of the valence band), or anisotropic (e.g. conduction bands at X or L), and present several phonon modes. 

In the present work, we address such issues. 
We keep the continuum hypothesis inherent to the large polaron \fro model, but waive the isotropic and non-degeneracy hypotheses, and also include multiple phonon branches. 
For polaron effective masses, working at the lowest order of perturbation theory, we provide analytical
results for the case of anisotropic electronic energy dispersion, with two distinct effective masses (uniaxial) and numerical simulations for the degenerate 3-band case, typical of III-V and II-VI semiconductor valence bands. 
We also deal with the strong-coupling limit, using a variational treatment: we propose trial wavefunctions for the above-mentioned cases, providing polaron radii and energies. Then, we evaluate the polaron formation energies,  effective masses and localisation lengths using parameters representative of a dozen II-VI, III-V and oxide semiconductors, for both electron and hole polarons. 
We show that for some cases perturbation theory  (weak-coupling approach) breaks down. 
In some other cases, the strong-coupling approach reveals that the large polaron hypothesis is not valid, which is another distinct breakdown.
In the non-degenerate case, we compare the perturbative approach with the Feynman path integral approach in characterisizing polarons in the weak coupling limit.
Thus, based on theoretical results for cubic materials, the present study characterize the validity of the continuum hypothesis for a large set of twenty materials.
\end{abstract}


\maketitle

\section{Introduction}
\label{sec:intro}

Polarons are usually defined as a correlated state combining a charge carrier and a deformation of the lattice. They are ubiquitous quasiparticles arising naturally in many insulators and semiconductors.\cite{Franchini2021} Although the polaron concept can be traced back to work by Landau\cite{Landau1933} in the thirties, the polaron term was introduced later by Pekar\cite{Pekar1946}. 
Early models assumed the localization length of the polaron to be much larger than the crystal atomic periodicity (large polaron), so that the charged particle couples with the polarization of the lattice through a macroscopically screened Coulomb interaction.
A formalism for determining large polaron properties based on a strong-coupling approach was provided first by Landau and Pekar.\cite{Landau1948} 
Later, \fro \cite{Frohlich1950} introduced a second-quantized formulation, and worked out a weak-coupling approach to large polarons. 
The \fro model can be studied also by more sophisticated formalisms, such as Feynman's path integral variational approach\cite{Feynman1955} or diagrammatic quantum Monte Carlo\cite{Mishchenko2000}, valid for the whole range of coupling strengths. 
Such approaches target polaron formation energy, effective mass, mobility, and optical characteristics
~\cite{Devreese2009}.
Alternatively, the Holstein model has been a successful theoretical tool for studying small polarons, in which the discrete character of the lattice is retained, and the electron-phonon interaction is short-ranged.\cite{Holstein1959a, Holstein1959b}

The full complexity of materials
is however not captured by these models. 
Most real materials have several phonon branches and electronic bands, both with arbitrary dispersions (albeit quadratic close to the extrema), and general electron-phonon interaction, though it generically reduces to the \fro form for small momentum transfer. Such features can only be addressed by first-principles approaches, that have seen an impressive development during the last two decades.
Existing first-principle approaches consider polarons either in the self-trapped adiabatic state~\cite{Stoneham2007, Varley2012, Sadigh2015, Geneste2017, Kokott2018, Franchini2021}, corresponding to the strong-coupling limit, or from many-body perturbation theory, corresponding
to the weak-coupling limit, e.g. the Allen-Heine-Cardona approach\cite{Marini2008, Giustino2010, Verdi2015, Ponce2015, Giustino2017, Sio2019, Miglio2020, Neaton2020}.
In the first case, the polaron formation energy is computed from the associated collective atomic displacement pattern and frozen electronic density, and the mobility is estimated through the computation of barriers for transitioning the localized polaron from one site to another.
In the second case, the treatment of the correlation between electron and phonon delivers a modification of the band edge eigenenergies, and a modification of the band gap often termed zero-point renormalization (ZPR), and also allows one to compute spectral functions.
Only recently a path towards a unified first-principles theory of polarons has been sketched~\cite{Sio2019}, and will likely be followed by further developments.

In Ref.~\onlinecite{Miglio2020} Miglio, Brousseau-Couture and coworkers established the predictive power of the first-principles approach for the ZPR,
in the Allen-Heine-Cardona methodology, by studying a set of 30 materials, and performed systematic comparison with existing experimental data. 
The inclusion of  non-adiabatic effects was demonstrated to be essential to produce a match between experimental and theoretical ZPR of the band gap. 
This was made clear
by introducing a generalized \fro model (gFr) from which the ZPR could be deduced without resorting to
full first-principles calculations, as only a few macroscopic quantities, determined either from experiment or first-principles calculations were needed.
This generalized \fro model will be at the center of the present work.

Indeed, the continuum hypothesis inherent to the \fro model can be dissociated from further simplifications present in the original \fro model. These additional simplifications are: (i) only one isotropic LO phonon branch, (ii) only one isotropic non-degenerate electronic band (iii) isotropic dielectric tensor to determine the electron-phonon coupling.
One can waive such restrictions without addressing the full complexity of the materials, in particular without introducing full-fledged electronic and phononic dispersions in the full Brillouin zone. 
By considering parabolic bands and dispersionless 
phonon branches, still allowing anisotropic behaviors and including non-degeneracy and warping effects~\cite{Mecholsky2014}, one captures an important
part of the electron-phonon effect in real materials, so that such generalized \fro model is sufficiently predictive for the most ionic materials among those considered, namely for the oxides and II-VI materials.
Miglio, Brousseau-Couture et al. provided the polaron formation energy within the lowest-order perturbation theory for the generalized \fro model. 
For each of the examined materials, less than a dozen parameters defined the gFr model, depending on the symmetries and number of atoms per cell. 

The original \fro model, that is characterized by one single adimensional parameter, $\alpha$, has been studied by many different theoretical approaches. 
These studies determined the ground state energy as a function of $\alpha$, but also the polaron effective mass, mobility, optical properties, localization, or modifications due to magnetic field, be they in tri-dimensional or confined situations.\cite{Devreese2007, Titantah2001, Hahn2018a}
The polaron effective mass, determined in the weak-coupling regime at the lowest order of perturbation, 
diverges when $\alpha$ is 6 or bigger, indicating the breakdown of perturbation theory. 
The localization radius, estimated from the strong-coupling limit, provides a criterion to determine whether the large polaron picture is correct: if it is comparable to or only slightly bigger than the distance between equivalent sites, the large polaron hypothesis breaks down.

So far, there have only been a few attempts to  formulate and study an extended model beyond the above-mentioned simplifying hypotheses. 
Trebin and R\"ossler~\cite{Trebin1975} studied polaron energies and effective masses in the case of triply degenerate bands, however without the inclusion of band warping (they worked with ``isotropic" triply degenerate bands), and 
ignored the effect of multiple LO phonon branches. 
Similarly, Fock, Kramer and B\"{u}ttner~\cite{Fock1975} examined polaron energies and masses in the non-degenerate case with uniaxial symmetry for effective mass (and dielectric tensor), but considered only one LO phonon branch. 
The case of multiple phonon branches was tackled by
Schlipf, Ponc{\'e}, Verdi and Giustino~\cite{Schlipf2018}, although they focused on the non-degenerate case with isotropic effective masses. 
Exact formulas were derived in these studies, while others~\cite{Mahan1965, Devreese2010} proposed approximate results.
To our knowledge, the strong-coupling limit has not been examined in any of these cases. 
A supplementary issue that is not considered in the present work has been recently tackled by Houtput and Tempere \cite{Tempere2021} in deriving anharmonic contributions to the \fro model.

In order to study polarons in more realistic cases, and pursue the work of Miglio, Brousseau-Couture et al., who focused on the polaron energies only, we compute effective masses in the weak-coupling limit, and further estimate 
localization lengths in the strong-coupling limit. 
We limit our investigation of the generalized \fro model to the case of cubic materials, treated in the harmonic approximation for phonons, and in the same macroscopic continuum hypothesis as in the usual \fro model.
We establish criteria for the breakdown of
perturbation theory, similar to that for a single isotropic non-degenerate electronic band, and obtain an estimation of the different (anisotropic) localization lengths, to be compared with the inter-atomic distance.

In Sec.~\ref{sec:theory_back} we summarize the basic theoretical knowledge of large polarons properties. First, we consider results for the standard \fro model, in the weak and strong coupling limits, then present the Feynman variational approach, valid for the whole coupling range.
We report the polaron formation energy from the gFr model, specialized for the
cubic case, for isotropic and anisotropic, non-degenerate and degenerate electronic dispersions.
In Sec.~\ref{sec:eff_mass} we derive the polaron effective mass 
for non-degenerate anisotropic as well as degenerate electronic bands coupled to multiple phonon modes. 
In Sec.~\ref{sec:localization} we provide the localization lengths beyond the basic \fro model, for the same scenarios as for the polaron effective masses. 
In Section \ref{sec:result}, we present our results (effective masses and localization lengths) 
for a subset of 20 cubic materials
among those originally considered by Miglio, Brousseau-Couture et al. 
In the discussion section \ref{sec:discussion},
we combine the effective masses and localization length data, to characterize, for each material and both electron and hole polarons, whether there is a breakdown of perturbation 
theory and whether the continuum hypothesis is valid or not. 
Such breakdowns happen indeed for several oxyde materials.
We also compare, for the non-degenerate uniaxial case, our formation energies and effective masses with those
from the Feynman path integral approach for the
isotropic case with effective masses equal to those of the uniaxial case, approximately averaged.
We conclude in Section \ref{conclusion}.

\section{Theoretical Background}
\label{sec:theory_back}

\subsection{Polaron energy, effective mass and localization radius in the \fro model}
\label{sec:theory_Fr}

The basic \fro model\cite{Frohlich1952} starts from one electronic band with spherical symmetry (i.e. isotropic), with parabolic dispersion determined by the effective mass $m^*$, coupled to one dispersionless longitudinal optical phonon of frequency $\omega_{LO}$ while ignoring the crystal discrete nature (continuum limit). 
The \fro model is thus able to describe a very reduced set of realistic scenarios, where e.g. compounds are binary,
the band extrema are neither anisotropic nor degenerate,
and one ignores the contribution to the polaron formation due to other phonon modes (e.g. acoustic, transverse optical). 

Let us first recall the \fro Hamiltonian describing this picture (everything will be expressed in atomic units unless stated otherwise: $m_e$ = 1, $|e|$ = 1, $\hbar$ = 1) :

\begin{equation}
    \label{eqn:FrH}
    \hat{H}^{Fr} = \hat{H}^{Fr}_{e} + \hat{H}^{Fr}_{p} + \hat{H}^{Fr}_{EPC},
\end{equation}

\noindent
where 

\begin{equation}
    \hat{H}^{Fr}_{e} = \sum_{\Vec{k}} \frac{\Vec{k}^2}{2m^*} \hat{c}^{+}_{\Vec{k}} \hat{c}^{\phantom{+}}_{\Vec{k}},
\end{equation}
 
\begin{equation}
    \hat{H}^{Fr}_{p} = \sum_{\Vec{q}} \omega_{LO} \hat{a}^{+}_{\Vec{q}} \hat{a}^{\phantom{+}}_{\Vec{q}}, 
\end{equation}

\begin{equation}
    \hat{H}^{Fr}_{EPC} = \sum_{\Vec{k},\Vec{q}} g^{Fr}(\Vec{q}) \hat{c}^{+}_{\Vec{k}} \hat{c}^{\phantom{+}}_{\Vec{k}} (\hat{a}^{\phantom{+}}_{\Vec{q}} + \hat{a}^{+}_{-\Vec{q}} ),
\end{equation}

\noindent
where $\hat{c}^{+}_{\Vec{k}}$($\hat{c}^{\phantom{+}}_{\Vec{k}}$) is the electron creation (annihilation) operator and $\hat{a}^{+}_{\Vec{q}}$ ($\hat{a}^{\phantom{+}}_{\Vec{q}}$) is the phonon creation (annihilation) operator. The electron-phonon coupling is defined as follows:

\begin{equation}
    g^{Fr}(\Vec{q}) = \frac{i}{q} \Bigg[ \frac{2\pi\omega_{LO}}{V_{BvK}} \Bigg( \frac{1}{\epsilon^{\infty}} - \frac{1}{\epsilon^{0}} \Bigg) \Bigg]^{1/2},
\end{equation}

\noindent
where $q$ is the norm of the phonon wave-vector $\Vec{q}$ and $V_{BvK}$ is the Born-von Karman volume.
This, in turn, can be rewritten as 

\begin{equation}
    g^{Fr}(\Vec{q}) = \frac{i}{q} \Bigg[ \frac{2\sqrt{2}\pi}{V_{BvK}} \frac{\omega_{LO}^{3/2}}{\sqrt{m^{*}}} \alpha \Bigg]^{1/2},
\end{equation}

\noindent
with the adimensional parameter $\alpha$ 

\begin{equation}
    \alpha = (\epsilon^*)^{\text{-}1} (m^*)^{1/2} \Big( 2\omega_{jLO} \Big)^{ {\text{-}1/2}}, 
\label{eq:alphaFr}
\end{equation}

\noindent
where 
\begin{eqnarray}
    \label{eqn:epsilon0_m1}
    (\epsilon)^{*\text{-}1}
    =
    (\epsilon^\infty)^{\text{-}1}
    -
    (\epsilon^{0})^{\text{-}1}.
\label{eq:dielconst}
\end{eqnarray}


The treatment of the \fro Hamiltonian at the lowest order of perturbation theory in $\alpha$ (second-order perturbation theory in the strength of the electron-phonon coupling) gives the polaron binding energy as a function of the crystalline momentum:

\begin{equation}
    E_{P}(\Vec{k}) = \frac{k^2}{2m^*} - \alpha 
    \frac{ \omega_{LO}}{k a_{LO} } \arcsin (k a_{LO}).
\end{equation}
where we define a characteristic length~\cite{Nery2016}
\begin{equation}
    a_{LO} = \Big( 2m^*\omega_{LO}\Big)^{1/2}.
\end{equation}
\noindent
Expanding the $\arcsin$ in a series up to the 2$^{nd}$ order term in $k a_{LO}$ one obtains the (isotropic) polaronic dispersion relation:

\begin{equation}
    E_{P}(\Vec{k}) = -\alpha\omega_{LO} + \frac{k^2}{2m^*}\Bigg( 1 - \frac{\alpha}{6}\Bigg) + \mathcal{O}(k^{4}),
\label{eq:epolaron_weak}
\end{equation}

\noindent
with the zero momentum value also known as the Zero Point Renormalization (ZPR).
\noindent
From the previous expression the polaron effective mass expression in the lowest-order perturbation theory is

\begin{equation}
    m^{*}_{P} = m^{*} \Bigg( 1 - \frac{\alpha}{6}\Bigg)^{-1}.
\label{eqn:spherical_effmass}
\end{equation}

This polaron effective mass expression changes sign for $\alpha \geq 6$, indicating the breakdown of perturbation theory. 
Even so, it is important to note that for values of the coupling constant close to 6 the polaron formation energy in the lowest-order perturbation theory treatment shows only mild deviations ($\sim$ 10-15\%) with respect to the diagrammatic Monte Carlo treatment.\cite{Mishchenko2000} 
Considering that we are dealing with lowest-order perturbation theory, the breakdown limit can be taken as a qualitative one, which signals a crossing from the weak to a strong coupling regime in the \fro picture. 
In this latter scenario the polaron reaches a self-trapping regime.

So far we have discussed the energetics of the polaron and effective mass. We now tackle another important characteristic of a polaron, namely, its localization length.
In the simplest case where the electronic band is isotropic, the theory describing the polaron localization has been developed by Pekar\cite{Pekar1946} and Landau and Pekar\cite{Landau1948} in the strong coupling theory framework, using a variational approach based on a Gaussian wavefunction ansatz. This method is described extensively in Mahan, section 7.1.3\cite{Mahan2000}. The approach is based on the variational principle, whereby the energy associated to a normalized trial wavefunction $\phi(\Vec{r})$ combines a electronic kinetic energy part and a coupled electron-phonon and phonon part. Explicitly,
\begin{equation}
E_P[\phi(\Vec{r})]=
E_{el}[\phi(\Vec{r})]+E_{epp}[\phi(\Vec{r})],
\label{eqn:E_P_var}
\end{equation}
where the electronic energy writes
\begin{equation}
E_{el}[\phi(\Vec{r})]=\int d\Vec{r}\frac{\big| \Vec{\nabla} \phi(\Vec{r}) \big|^2}{2m^*},
\label{eqn:E_el_iso}
\end{equation}
while the electronic density,
\begin{equation}
\rho(\Vec{r})=|\phi(\Vec{r}) \big|^2,
\label{eqn:rho_r}
\end{equation}
induces a response of the polarizable medium,
obtained mathematically by considering its
Fourier transform
\begin{equation}
\rho(\Vec{k})=\frac{1}{(2\pi)^{3/2}}
\int d\Vec{r} \rho(\Vec{r}) e^{i \Vec{k} \Vec{r}}
\label{eqn:rho_k}
\end{equation}
and building the associated electron-phonon and phonon energy
\begin{equation}
E_{epp}[\phi(\Vec{r})]=-\int d\Vec{k}
\frac{4 \pi}{2k^2\epsilon^{*}}
\rho^*(\Vec{k}) \rho(\Vec{k})
.
\label{eqn:E_epp_iso}
\end{equation}

The electron wavefunction is frozen, as well as the deformation of the polarizable medium. 
Hence, in this strong-coupling scenario, one works in an adiabatic approximation, in which the electron and phonon parts are explicitly correlated, albeit frozen. 
A discussion of the leftover part is also provided in Ref.~\onlinecite{Mahan2000}, showing how the variational approach was obtained from the full second-quantized Hamiltonian, to which both weak- and strong-coupling approaches can be applied, as well as the other methodologies mentioned in the introduction.

Different simple trial wavefunctions can be found in the literature, giving analytical results.  
The convenient Gaussian trial wavefunction\cite{Mahan2000} writes
\begin{equation}
    \phi(\Vec{r}) = \Bigg( \frac{1}{\sqrt{\pi}a_P} \Bigg)^{3/2}\exp{\Bigg( -\frac{r^{2}}{2a_P^2} \Bigg)},
\end{equation}

\noindent
where $a_P$ is the localization radius of the electron, an adjustable parameter in the polaron wavefunction, allowing the minimization of the polaron formation energy.

The best (in the variational sense) localization length is then obtained in terms of the electron effective mass $m^{*}$ and the dielectric constant $\epsilon^{*}$, defined in Eq.~(\ref{eq:dielconst}): 
\begin{equation}
    \label{eqn:ap_sphere}
    a_{P} = \frac{3}{2} \sqrt{2\pi} (m^*)^{\text{-}1} \epsilon^{*}.
\end{equation}
\noindent
This adiabatic localization length is independent on the phonon frequency, and is roughly 17$\%$ larger than the polaron localization length derived using an exponential wavefunction ansatz (see Sio et al. \cite{Sio2019}, Eq. 11).
The polaron localization can be expressed in terms of the phonon energy if one reintroduces the $\alpha$ parameter:
\begin{equation}
    a_{P} = \frac{3}{2} \sqrt{\pi}
    (m^{*}\omega_{LO})^{-1/2} (\alpha)^{-1}.
\end{equation}
The localization length is not a hard limit on the polaron extent, but indicates a characteristic decay length of the self-trapped electronic wavefunction.

Along the same lines one can provide the polaron ground state energy in the spherical limit within the Gaussian variational strong-coupling approach:

\begin{equation}
    \label{eqn:e_sphere}
    E_{P} = -\frac{1}{6\pi}m^{*}(\epsilon^{*\text{-}1})^{2}.
\end{equation}

\noindent
This Gaussian polaron stabilization energy is roughly 8$\%$ lower than with the exponential ansatz (see Sio et al. \cite{Sio2019}, Eq. 12), suggesting
a Gaussian wavefunction ansatz seems to be a superior choice to start with in the variational approach.

Rewriting the previous equation in term of the coupling constant $\alpha$, one arrives at:

\begin{equation}
    E_{P} = -\frac{1}{3\pi}\omega_{LO}\alpha^{2}
    \approx -\omega_{LO} ( 0.1061 \alpha^{2}).
\label{eq:epolaron_strong}    
\end{equation}

A much more sophisticated attempt in determining the polaron ground state energy in the strong-coupling limit has been provided by Miyake\cite{Miyake1976}, which shows a slightly different prefactor
of the term quadratic in the coupling constant $\alpha$, and adds a constant term, in a perturbation expansion with small parameter $1/\alpha^2$:

\begin{equation}
    E_{P} \approx -\omega_{LO} ( 0.1085 \alpha^{2} + 2.836 + \mathcal{O} (1/\alpha^2) ).
\end{equation}

In the weak-coupling regime the polaron ground state energy shows a leading linear dependence on $\alpha$, while in the strong coupling regime a quadratic one. 
The crossing between the two different regimes, weak and strong, given by Eqs.~(\ref{eq:epolaron_weak}) and  (\ref{eq:epolaron_strong}) gives a value of $\alpha \approx 9.5$ (leaving aside the constant shift in the strong-coupling expansion). 

\subsection{Polaron energy, effective mass and radius in the Feynman approach}
\label{sec:Feynman_model}

The Feynman \cite{Feynman1955} 1955 path integral approach to the \fro model casts the \fro Hamiltonian (Eq. (\ref{eqn:FrH})) into a Lagrangian by passing the electron and phonon creation and annihilation operators into the corresponding coordinates and momenta, and performing a Legendre transformation. 
After a Gaussian integration over the momenta, one is left with a configuration path integral that is Gaussian over the phonon coordinates which can be evaluated directly. 
This results in an effective action of an electron coupled by a non-local Coulomb potential to a second fictitious particle:
\begin{equation}
    \label{eqn:feynman_pi}
    Z = \textrm{Tr} \exp{(-\beta\hat{H}^{Fr})} \approx \int_{\Vec{r}(0) = \Vec{r'}(\beta)} \mathcal{D}\Vec{r}(\tau)\ e^{-S[\Vec{r}(\tau)]},
\end{equation}
with
\begin{equation}
\begin{gathered}
    \label{eqn:model_action}
    S[\Vec{r}(\tau)] = \frac{m^*}{2} \int^{\beta}_{0} d\tau \left( \frac{d \Vec{r}(\tau)}{d\tau} \right)^2 - \frac{( \omega_{LO})^{3/2} \alpha}{\sqrt{8 m^*}} \\
    \times \int^{\beta}_0 d\tau \int^{\beta}_0 d\sigma |\Vec{r}(\tau) - \Vec{r}(\sigma)|^{-1} e^{-\omega_{LO}|\tau - \sigma|}.
\end{gathered}
\end{equation}
Here $S$ is the (\fro) \emph{model} action, and $Z$ is the partition function for the electron to begin and end at the same position over imaginary time $i\hbar\beta$, with the thermodynamic beta $\beta = 1/k_B T$. 
This form of $S$ already assumes that $\beta$ is large (low temperature, relative to the phonon energy).
\newline

The path integral is difficult to evaluate for the $1/r$ Coulomb potential, so Jensen's inequality, $\langle \exp{f} \rangle \geq \exp{\langle f \rangle}$, is used to approximate the model action $S$ by an analytically path-integrable \emph{quadratic} action (the \emph{trial} action), $S_0$, that has a time-retarded potential and a (tunable) exponential factor similar to $S$:
\begin{equation}
\begin{gathered}
    \label{eqn:trial_action}
    S_0[\Vec{r}(\tau)] = \frac{m^*}{2} \int^{\beta}_{0} d\tau \left( \frac{d \Vec{r}(\tau)}{d\tau} \right)^2 \\
    + \frac{C}{2} \int^{\beta}_0 d\tau \int^{\beta}_0 d\sigma \left(\Vec{r}(\tau) - \Vec{r}(\sigma)\right)^{2} e^{-w |\tau - \sigma|}.
\end{gathered}
\end{equation}
This results in the Feynman-Jensen inequality that gives a solvable upper-bound to the model free energy,
\begin{equation}
    \label{eqn:feynman_jensen}
    F \leq F_{S_0} + \frac{1}{\beta} \langle S - S_0 \rangle_{S_0} , \\
\end{equation}
where $F_{S_0}$ is the free energy of the trial system and $\langle S - S_0 \rangle_{S_0}$ is the expectation value of the difference in the two actions, taken with respect to the trial system,
\begin{equation}
    \label{eqn:pi_avg}
    \langle S - S_0 \rangle_{S_0} = \frac{\int \mathcal{D}\Vec{r}(\tau)(S - S_0)e^{-S_0[\Vec{r}]}}{\int \mathcal{D}\Vec{r}(\tau)e^{-S_0[\Vec{r}]}}.
\end{equation}
The ground-state energy $E$ is estimated by taking the limit $\beta \rightarrow \infty$ (temperature goes to zero). 
In this limit the trace in Eq. (\ref{eqn:feynman_pi}) can be estimated by the ground-state, $\textrm{Tr} \exp(-\beta \hat{H}) |_{\beta \to \infty} \approx \exp(-\beta E)$. 
The Feynman-Jensen inequality for the ground-state energy is thus:
\begin{equation}
\begin{gathered}
    \label{eqn:gsenergy_feynman}
    E \leq  \frac{3 \omega_{LO}}{4 v} \left(v-w\right)^2 - \frac{\alpha \omega_{LO}}{\sqrt{\pi}} \frac{v}{w} \int^\infty_0 d\tau \frac{e^{-\tau}}{\sqrt{\tau D(\tau)}} , \\ 
    D(\tau) = 1 + \frac{v\left(1 - w^2 / v^2\right) \left(1 - e^{-v \tau}\right)}{\tau w^2}, 
\end{gathered}
\end{equation}
where $v^2 \equiv w^2 + 4C/w$. 
The process is variational with two free parameters: $v$ (the frequency of the harmonic oscillator composed of the electron and the fictitious particle), and $w$ (the exponential rate of decay of the coupling in time). 

For extremal values of the coupling $\alpha$, $w$ smoothly approaches limits of $1$ and $3$. 
In the weak-coupling (small alpha) limit the energy minimum occurs when $v$ is near $w$. Therefore, Feynman set $v = (1 + \varepsilon) w$ where $\varepsilon$ is small and expanded the energy expression (RHS of Eq. (\ref{eqn:gsenergy_feynman})) w.r.t $\varepsilon$. Feynman then minimised the energy first w.r.t $\varepsilon$ and then w.r.t $w$ and found that in the weak-coupling limit the energy is least when
\begin{equation}
\begin{gathered}
    \label{eqn:weak_gs_feynman}
    \frac{w}{\omega_{LO}} = 3, \quad \frac{v}{\omega_{LO}} = 3 \left[ 1 + \frac{2\alpha}{3w} \left(1 -\frac{2}{w} \left[ \sqrt{w-1}-1 \right] \right) \right] ,\\
    \frac{E}{\omega_{LO}} \leq -\alpha - 1.23 \left(\frac{\alpha}{10}\right)^2 .
\end{gathered}
\end{equation}

In the strong-coupling (large alpha) limit $v$ is large and $w$ approaches 1 so $w / v << 1$. Therefore, Feynman expanded the energy expression w.r.t $w / v$ and then minimised the energy w.r.t $v$ and $w$ and found that in the strong-coupling limit the energy is least when 
\begin{equation}
\begin{gathered}
    \label{eqn:strong_gs_feynman}
    \frac{w}{\omega_{LO}} = 1, \quad \frac{v}{\omega_{LO}} = \frac{4\alpha^2}{9\pi} - 4\left( \log2 + \frac{1}{2} \gamma \right) + 1 , \\
    \frac{E}{\omega_{LO}} \leq -\frac{\alpha^2}{2\pi} - \frac{3}{2}(2\log2+\gamma)-\frac{3}{4} + \mathcal{O}\left(\frac{1}{\alpha^2}\right) . \\
\end{gathered}
\end{equation}
where $\gamma = 0.5772 \dots$ is the Euler-Mascheroni constant. 

For finite temperature a numeric variational solution is required. As temperature is relative to the phonon energy, soft polar materials will require a numeric solution at room temperature. The variational parameters for finite temperature can be evaluated by numerically minimising \=Osaka's finite temperature generalisation of Feynman's variational principle \cite{Osaka1959} to give a lower upper-bound to the polaron free energy.

The effective polaron mass at zero temperature was found by Feynman\cite{Feynman1955} by assuming that the electron moves with a small velocity $\Vec{u}$ from an initial coordinate $\Vec{0}$ to a final coordinate $\Vec{r}_f = \Vec{u} \hbar\beta$ in an imaginary-time $\hbar\beta$. Feynman then sought the total energy of the polaron and equated it to the form $E_0 + \frac{1}{2}m^*_P u^2$ by expanding the total energy expression to quadratic-order in the velocity $\Vec{u}$. From the kinetic energy term Feynman found the polaron effective mass:
\begin{equation}
\begin{gathered}
    \label{eqn:mass_feynman}
    m^{*F}_P = m^* \left[ 1 + \frac{\alpha}{3\sqrt{\pi}} \left(\frac{v}{w}\right)^3 \int^\infty_0 d\tau \frac{e^{-\tau} \tau^{1/2}}{\left[D(\tau)\right]^{3/2}} \right].
\end{gathered}
\end{equation}
where the values of the variational parameters are those that minimise the original polaron ground-state energy when $u = 0$ in Eq. (\ref{eqn:gsenergy_feynman}). From the values in Eq. (\ref{eqn:weak_gs_feynman}) Feynman obtained the weak-coupling expression 
\begin{equation}
    \label{eqn:weak_mass_feynman}
    m^{*F}_P = m^* \left[ 1 + \frac{1}{6} \alpha + 0.025 \alpha^2 + \dots \right] ,
\end{equation}
and from the values in (\ref{eqn:strong_gs_feynman}) the strong-coupling expression is
\begin{equation}
    \label{eqn:strong_mass_feynman}
    m^{*F}_P = m^* \frac{160}{81} \left(\frac{\alpha}{\pi}\right)^4.
\end{equation}
At finite temperatures the effective polaron mass is proportional to the imaginary part of the complex impedance function $Z(\nu, \beta)$ provided by Feynman et al. \cite{Feynman1962} (Eqs. (35), (36) \& (41)) in the zero frequency limit $\nu \to 0$:
\begin{equation}
    m^{*F}_P(\beta) =  m^* \lim_{\nu \to 0} \left\{ \frac{\text{Im} Z(\nu, \beta)}{\nu} \right\}
\end{equation}
as described by Peeters and Devreese \cite{Peeters1984}.

Schultz \cite{Schultz1959} estimated the polaron size by calculating the root mean square distance between the electron and the fictitious particle. 
The reduced mass of their relative motion is:
\begin{equation}
    \label{eqn:red_mass_schultz}
    \mu = \frac{m^*}{m^{*F}_P} (m^{*F}_P - m^*).
\end{equation}
Schultz then used the ground-state harmonic oscillator wave function for the relative coordinate $\rho$,
\begin{equation}
    \label{eqn:gs_wavefunc_schultz}
    \phi_0(\rho) = \left( \frac{\mu v \omega_{LO}}{\pi} \right)^{3/4} \exp\left( -\frac{\mu v \omega_{LO} \rho^2}{2 } \right),
\end{equation}
to define a polaron radius:
\begin{equation}
    \label{eqn:pol_size_schultz}
    r_P \equiv \langle \rho^2 \rangle^{1/2} = \frac{1}{2} \left(\frac{3}{\mu v \omega_{LO}}\right)^{1/2}.
\end{equation}
In the weak-coupling limit this is
\begin{equation}
    \label{eqn:weak_size_schultz}
    r_P \sim \frac{3}{4} \left(\frac{6}{\alpha m^* \omega_{LO}}\right)^{1/2},
\end{equation}
and in the strong-coupling limit 
\begin{equation}
    \label{eqn:strong_size_schultz}
    r_P \sim \frac{3 }{2\alpha} \left(\frac{\pi}{m^* \omega_{LO}}\right)^{1/2}.
\end{equation}

In the degenerate anisotropic uniaxial case, we propose to na\"ively incorporate the anisotropy into the Feynman approach (which is one-dimensional due to the underlying isotropy of the \fro Hamiltonian) by treating the two directions independently with effective masses $m_\perp$ and $m_z$. 
We then use the variational principle separately in each direction to find the variational parameters $v_{\perp/z}$ and $w_{\perp/z}$ that give the lowest upper-bound to the ground-state energy $E_{\perp/z}$ for each direction. 
The variational parameters can then be used to obtain the effective polaron masses $m^{*F}_{P,\perp}$ and $m^{*F}_{P, z}$ using Eq. (\ref{eqn:mass_feynman}) and polaron radii $r_{P\perp}$ and $r_{Pz}$ using Eq. (\ref{eqn:pol_size_schultz}). 

We define an effective ground-state energy by taking the arithmetic mean, 
\begin{equation}
    E = \frac{2 E_\perp + E_z}{3} ,
\end{equation}
and the effective radius of the anisotropic polaron by the geometric mean, 
\begin{equation}
    r_P = \left(r^2_{P\perp} r_{Pz}\right)^{1/3}.
\end{equation}







\subsection{Polaron formation energy in the generalized \fro cubic model}
\label{sec:energy_gFr}

The generalized \fro model is considered now in the specific case of cubic materials. The most general case of the \fro model without restriction on the crystallographic systems is discussed elsewhere.\cite{Miglio2020} 
In particular the cubic case arrives with certain advantages: the macroscopic dielectric tensor $\epsilon^{\infty}$ and, in the neighborhood of the zone-center, the phonon dispersion do not depend on the wavevector direction.
The results are now reformulated 
to simplify their structure.

Considering both the electron polaron and the hole polaron in cubic materials (corresponding to conduction states and valence states), the variable $\sigma$, that takes value $\sigma=1$ for electron polarons and $\sigma=-1$ for hole polaron, is introduced to characterize the electronic band dispersion, which will be used to ensure the effective mass is always positive at a band extremum.

The gFr Hamiltonian in the cubic case is expressed as follows:

\begin{equation}
\begin{split}
    \hat{H}^{gFr} = \sum_{\Vec{k},n} \frac{\sigma \Vec{k}^2}{2m^{*}_{n}(\hat{k})} \hat{c}^{+}_{\Vec{k},n} \hat{c}^{\phantom{+}}_{\Vec{k},n} + \sum_{\Vec{q},j} \omega_{jLO} \hat{a}^{+}_{\Vec{q},j} \hat{a}^{\phantom{+}}_{\Vec{q},j} \\
    + \sum_{\Vec{k}nn',\Vec{q}j} g^{gFr}(\Vec{k}nn',\Vec{q}j) \hat{c}^{+}_{\Vec{k}+\Vec{q},n'} \hat{c}^{\phantom{+}}_{\Vec{k},n} (\hat{a}^{\phantom{+}}_{\Vec{q},j} + \hat{a}^{+}_{-\Vec{q},j} ).
\end{split}
\label{eq:cfr}
\end{equation}

One can point out several aspects in Eq. (\ref{eq:cfr}), with respect to the usual \fro Hamiltonian.  The electronic effective mass $\hat{k}$-dependence in this case accounts for possible anisotropic behaviour of the electronic dispersion. $\hat{k}$ is the normed direction of the $\Vec{k}$ vector. Index $n$ accounts for the summation over the electronic bands, thus one can consider possible degenerate states. The summation over $j$ includes the possibility to couple to multiple phonon branches, still considering only the $LO$ type. Most importantly, compared to the basic one-band model, the EPC term summation over the $n$ and $n'$ indices now includes interband contribution, if the bands are degenerate at the edge.

The EPC is given by:

\begin{equation}
\begin{split}
    g^{gFr}(\Vec{k}nn',\Vec{q}j) = \frac{i}{q} \frac{4\pi}{\Omega_{0}} \Bigg( \frac{1}{ 2\omega_{jLO} V_{BvK}}  \Bigg)^{1/2} \frac{p_{jLO}}{\epsilon^{\infty}} \\
    \times \sum_{m} s_{n'm}(\hat{k}')(s_{nm}(\hat{k}))^{*}
\end{split}
\end{equation}

\noindent
with $p_{jLO}$ the phonon mode-polarity and the $s(\hat{k})$ tensors are unitary matrices,\cite{Miglio2020} actually the overlap matrices between a fixed basis of wavefunctions at $\Gamma$ and the wavefunctions tending to $\Gamma$ along direction $\hat{k}$. The wavevector $\vec{k}'= \vec{k}+\vec{q}$ has direction $\hat{k}'$. In the non-degenerate case, $s$ is 1, irrespective of $\hat{k}$.

The polaron binding energy has been derived within the gFr model,\cite{Miglio2020} taking into account possible wavevector dependence of the phonon frequencies and mode-polarity vector, as well as anisotropic dielectric tensor. 
These do not show up in the cubic case, thus
Eq.(6) of Ref.~\onlinecite{Miglio2020}
reduces to:

\begin{eqnarray}
{\rm ZPR}^{\rm gFr}&=&
-\sum_{jn} 
\frac{\sigma}{\sqrt{2}\Omega_0 
n_{\rm deg}}
\Big(\int_{4\pi} d\hat{k}
\big( m_n^*(\hat{k})
\big)^{1/2}\Big)
\nonumber
\\
&\times &\big(\omega_{jLO}
\big)^{-3/2}
\Big(
\frac{p_{jLO}}
        {\epsilon^\infty}
\Big)^2.
\label{eq:ZPR_c_Fr}
\end{eqnarray}

In view of further developments, and in order to highlight the similarity with 
Eqs.~(\ref{eq:alphaFr}) and (\ref{eq:epolaron_weak}),
we rewrite this expression, and introduce the phonon branch-dependent generalization of the adimensional parameter $\alpha$:
\begin{equation}
    \label{eqn:alpha_cube}
    \alpha_{j} = < m^{*1/2} > \Big( 2\omega_{jLO} \Big)^{ {\text{-}1/2} }(\epsilon_{j}^*)^{\text{-}1},
\end{equation}

\noindent
where the average square-root effective mass is 
\begin{eqnarray}
    \label{eqn:av_effmass_sqr}
    < m^{*1/2} > = 
    \sum_{n}
    \frac{1}{n_{\rm deg}}
    \Big(\frac{1}{4\pi}\int_{4\pi}
    d\hat{k}
    \big(m^{*}_n(\hat{k})\big)^{1/2}\Big),
\end{eqnarray}
and the phonon branch $j$ contribution to
$\epsilon^{*\text{-}1}$ is
\begin{eqnarray}
    \label{eqn:epsilonj_m1}
    \epsilon_{j}^{*\text{-}1}=
    \frac{4\pi}{\Omega_0}
    \Big(
    \frac{p_{jLO}}
        {\omega_{jLO} \epsilon^\infty}
    \Big)^2.
\end{eqnarray}

One can check that the static dielectric constant is the sum of the electronic (optical) one and those
of all phonon branches~\cite{Gonze1997, Miglio2020}:
\begin{eqnarray}
    \label{eqn:epsilon0_m1_sumj}
    (\epsilon^{0})^{\text{-}1}=
    (\epsilon^\infty)^{\text{-}1}
    +
    \sum_{j}(\epsilon_{j}^*)^{\text{-}1}.
\end{eqnarray}

Such formulas generalize easily to the non-cubic case, but the cubic case allows a decoupling of the electronic directionality average from the vibrational and dielectric behaviour.

In term of these quantities, the polaron binding energy in the generalized cubic \fro model is expressed as: 

\begin{equation}
\begin{split}
    \text{ZPR} &= - \sigma \sum_{j} \alpha_{j} \omega_{jLO} \\
               &= - \sigma < m^{*1/2} > \sum_{j} \Big( 2\omega_{jLO} \Big)^{ {\text{-}1/2} }(\epsilon_{j}^*)^{\text{-}1}\omega_{jLO} .
\end{split}
\label{eqn:zpr_cube}
\end{equation}
In what follows, the vibrational and dielectric factor in this expression, as well as a similar expression without the rightmost factor $\omega_{jLO}$ will be encountered often, so that we define, for $n=0$ and $n=1$,
\begin{equation}
V_{n} = \sum_{j} \Big( 2\omega_{jLO} \Big)^{ {\text{-}1/2} }(\epsilon_{j}^*)^{\text{-}1}(\omega_{jLO})^n,
\label{eqn:V_n}
\end{equation}
which gives
\begin{equation}
    \text{ZPR} = - \sigma < m^{*1/2} > V_1 .
\label{eqn:zpr_short}
\end{equation}

\section{Polaron effective mass in cubic materials}
\label{sec:eff_mass}

We now derive the expressions for the polaron effective mass, in the weak-coupling limit, from many-body perturbation theory. The self-energy is obtained, then derived twice with respect to the wavevector, giving the effective mass. We will treat first the non-degenerate case, then the degenerate case.

\subsection{Effective masses in the non-degenerate case}
\label{sec:non_deg_eff_mass}

In order to determine the polaron effective mass, the self-energy $\Sigma(\Vec{k})$ is determined at the lowest order in many-body perturbation theory, in the on-the-mass-shell approach. 
The polaron energy $E_{P}(\Vec{k})$ is obtained as
\begin{equation}
    E_{P}(\Vec{k}) = E(\Vec{k}) + \Sigma(\Vec{k},E(\Vec{k})),
    \label{eqn:E_P_k}
\end{equation}
\noindent
with 
\begin{equation}
    \Sigma(\Vec{k},z) = \frac{1}{\pi} \sum_{j}
                       \int d^{3}q \frac{\omega_{jLO}}{4\pi q^{2}\epsilon^{*}_{j}} \frac{1}{ ( z - E(\Vec{k} + \Vec{q}) ) - \sigma\omega_{jLO} },
    \label{eqn:Sigma_z_k}
\end{equation}
where $E(\Vec{k})$ is the bare electron/hole energy depending on the wavevector $\Vec{k}$.

In the non-degenerate case, the bare electronic energy is expressed in terms of effective masses along the three principal axes of the electronic dispersion relation (with the general case of a band edge anywhere in the Brillouin Zone, note that cross terms like $m_{xy}$ can always be eliminated by the proper choice of axes).
Deviations from quadratic behavior are ignored, and are beyond the gFr model.
Thus
\begin{equation}
    E (\Vec{k} ) = \frac{\sigma}{2} \Bigg( \frac{k^{2}_{x}}{m^{*}_{x}} + \frac{k^{2}_{y}}{m^{*}_{y}} + \frac{k^{2}_{z}}{m^{*}_{z}} \Bigg).
        \label{eqn:E_k_effmass}
\end{equation}

After expanding the self-energy in a Taylor series up to second order in the wavevector components, the polaron dispersion energy in the non-degenerate energy case becomes 

\begin{equation}
    E_{P} (\Vec{k} ) \approx \Sigma(0,\Vec{k} = 0 ) + \frac{\sigma}{2} \Bigg( \frac{k^{2}_{x}}{m^{*}_{P,x}} + \frac{k^{2}_{y}}{m^{*}_{P,y}} + \frac{k^{2}_{z}}{m^{*}_{P,z}} \Bigg),
\label{eqn:E_P_k_effmass}
\end{equation}
\noindent
where $\Sigma(0,\Vec{k} = 0 )$ is the ZPR energy.
The quantities $m^{*}_{P,i}$ are the sought polaron effective masses:
\begin{equation}
    \frac{1}{m^{*}_{P,i}} = \frac{1}{m^{*}_i} + \sigma\frac{ d^{2} \Sigma(\Vec{k},E(\Vec{k}))}{ dk^{2}_{i}} \Big|_{k=0} ~;~ i = x,y,z .
    \label{eqn:m_P}
\end{equation}

Eqs.~(\ref{eqn:E_P_k}-\ref{eqn:m_P}) can be worked out as follows. One defines
\begin{equation}
    I_{P} (\Vec{k},\omega_{jLO},m^{*}_{i}) = 
    \int \frac{d^{3}q}{4\pi q^{2}} 
    \frac{1}{ \sigma( E(\Vec{k}) - E(\Vec{k} + \Vec{q}) ) - \omega_{jLO} },
\end{equation}
with the dependence of $I_P$ on effective masses being given by 
Eq.~(\ref{eqn:E_k_effmass}),
so that
\begin{equation}
    \Sigma(\Vec{k},E(\Vec{k} )) = \frac{\sigma}{\pi} \sum_{j} \frac{\omega_{jLO}}{\epsilon^{*}_{j}}
  I_{P} (\Vec{k},\omega_{jLO},m^{*}_{i}).
\end{equation}
One easily proves that
\begin{equation}
    I_{P} (0,\omega_{jLO},m^{*}_{i}) = 
    (\omega_{jLO})^{-1/2}I_{P} (0,1,m^{*}_{i})
\end{equation}
and
\begin{equation}
    \frac{ \partial^{2} I_P}{ \partial k^{2}_{i}} \Big|_{k=0}
    =
    -2\frac{ \partial^{2} I_P}{ \partial \omega_{jLO} \partial m^*_{i}} \Big|_{k=0,\omega_{jLO}},
\end{equation}
giving
\begin{equation}
\begin{split}
    \frac{ d^{2} \Sigma(\Vec{k},E(\Vec{k}))}{ dk^{2}_{i}} \Big|_{k=0} 
=
    \frac{\sigma}{\pi}
    \frac{\partial I_P}{\partial m^*_{i}} \Big|_{k=0,\omega=1}
    V_0,
\end{split}
\end{equation}
to be inserted in Eq.~(\ref{eqn:m_P}).
In this expression, the decomposition between dependence on the electronic dispersion, through the effective masses, and the dependence on the phonon branches is clear. 

The required expression for $I_P(0,1,m^{*}_{i})$, namely
\begin{equation}
    I_{P} (0,1,m^{*}_{i}) = 
    -\int \frac{d^{3}q}{4\pi q^{2}} 
    \frac{1}{ 
    \frac{1}{2} \Big( \frac{q^{2}_{x}}{m^{*}_{x}} + \frac{q^{2}_{y}}{m^{*}_{y}} + \frac{q^{2}_{z}}{m^{*}_{z}} \Big)
    +1 },
\end{equation}
relates directly to the average square root effective mass,
Eq.~(\ref{eqn:av_effmass_sqr}), in this non-degenerate case.
Indeed, $\vec{q}$ can be decomposed in its norm $q$ and direction $\hat{q}$, $\vec{q}=q \hat{q}$, the radial and angular integrations can be decoupled, and the radial one performed exactly. The effective mass in direction $\hat{q}$ is given by
\begin{equation}
    \frac{1}{m^{*}(\hat{q})} = 
    \frac{\hat{q}^{2}_{x}}{m^{*}_{x}} + \frac{\hat{q}^{2}_{y}}{m^{*}_{y}} + \frac{\hat{q}^{2}_{z}}{m^{*}_{z}},
\end{equation}
and one obtains
\begin{equation}
    I_{P} (0,1,m^{*}_{i}) = 
    -\frac{\pi}{\sqrt{2}}< m^{*1/2} >.
\end{equation}
Finally, one recovers first Eq.~(\ref{eqn:zpr_cube}), as ZPR$=\Sigma(\Vec{0},E(\Vec{0}))$, but one finds also
\begin{equation}
    \frac{1}{m^{*}_{P,i}} = \frac{1}{m^{*}_i} 
    -
    \frac{\partial < m^{*1/2} >}{\partial m^*_{i}} V_0.
    \label{eqn:m_P_from_average}
\end{equation}

In the non-degenerate isotropic case, e.g. when one has a single electronic band edge at $\Gamma$, the polaron effective mass simplifies to Eq. (\ref{eqn:spherical_effmass}). 

Moreover, one can treat also analytically the non-degenerate anisotropic uniaxial case, suitable to describe materials where the band extremum is found at $X$ or $L$ points in cubic materials, with a uniaxial electronic dispersion relation.
In this case, we define 
\begin{equation}
    m_{x} = m_{y} = m_{\perp},
    \label{eqn:m_uniaxial}
\end{equation}
as well as 
the ratio between  effective masses
$\mu^{*} = \frac{m_{\perp}}{m_{z}}$.
We can distinguish two possible scenarios, $\mu^{*} > 1$ and
$\mu^{*} < 1$.
The self-energy will inherit the uniaxial symmetry of the electronic structure and we can determine the in-plane and out-of-plane derivatives based solely on the two different electronic effective masses. 
Mathematically, one obtains the following formulas. 
The average square root of the effective mass
over all directions becomes
\begin{equation}
    <m^{*1/2}> = m^{1/2}_{\perp} S(\mu^{*} - 1).
\end{equation}
where
\begin{equation}
    \label{eqn:sfunction}
    S(x) = \begin{cases}
                    \frac{argsh (x^{1/2})}{x^{1/2}}, \, x > 0, \\\\
                    \frac{argsin [(-x)^{1/2}]}{(-x)^{1/2}}, \, x < 0,
                \end{cases}
\end{equation}
This yields analytical formulas for the electron-phonon modification of the inverse effective masses,

\begin{equation}
\begin{split}
    \frac{ d^{2} }{ dk^{2}_{\perp}} \Sigma \Bigg|_{k=0}  =  & \sigma \frac{1}{4\sqrt{2}} m^{-1/2}_{\perp}  \Bigg[ \frac{  S(\mu^{*}-1) - (\mu^{*})^{1/2}  }{ \mu^{*} - 1  }  \Bigg] V_0, 
\end{split}
    \label{eq:ds_dperp_uni}
\end{equation}
\noindent
and
\begin{equation}
    \frac{ d^{2} }{ dk^{2}_{z}} \Sigma \Bigg|_{k=0} = \sigma \frac{1}{2\sqrt{2}} m^{-1/2}_{z} \frac{ \mu^{*} }{ \mu^{*} - 1 } \Big[  1 - \mu^{*1/2} S(\mu^{*}-1)  \Big] V_0 .
    \label{eq:ds_dz_uni}
\end{equation}

\begin{figure}
    \centering
    \includegraphics[width=0.49\textwidth]{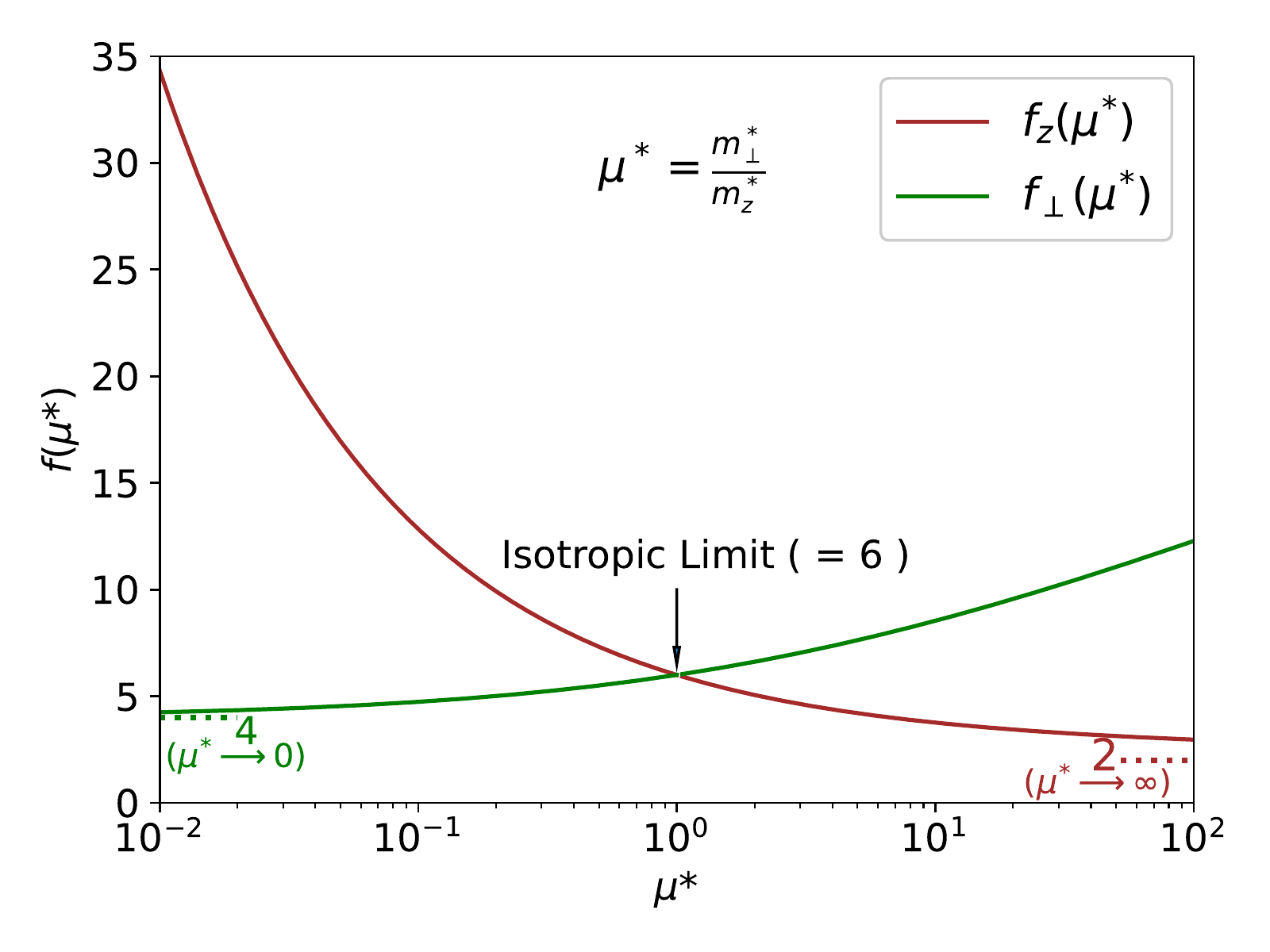}
    \caption{Polaron effective mass breakdown limit in the uniaxial case. Note that when the two relevant effective massess, $m^{*}_{z}$, respectively $m^{*}_{\perp}$, are equal we reach the isotropic limit breakdown of 6.}
    \label{fig:alpha_breakdown_uniaxial}
\end{figure}

The polaron effective mass is shown to reach the perturbation theory breakdown limit at a different rate when compared to the isotropic case, as it depends on the direction in the uniaxial case. The self-energy behaves differently along $\perp$- and $z$- directions, as shown in Eqs. (\ref{eq:ds_dperp_uni}) and (\ref{eq:ds_dz_uni}), and the polaron effective mass has the same behaviour:

\begin{equation}
    m^{*}_{P,\perp} = m^{*}_{\perp} \Big ( 1 - \frac{1}{f_{\perp}(\mu^{*})} \Sigma_{i}\alpha_{i} \Big)^{-1}
\end{equation}

\noindent
and

\begin{equation}
    m^{*}_{P,z} = m^{*}_{z} \Big ( 1 - \frac{1}{f_{z}(\mu^{*})} \Sigma_{i}\alpha_{i} \Big)^{-1};
\end{equation}

\noindent
with

\begin{equation}
    f_{\perp}(\mu^{*}) = 4 ( 1 - \mu^{*} ) \Big[  1 - \frac{(\mu^{*})^{1/2}}{S(\mu^{*} - 1)}  \Big]^{-1}
\end{equation}

\noindent
and

\begin{equation}
    f_{z}(\mu^{*}) = 2 \frac{( 1 - \mu^{*} ) }{\mu^{*} } \Big[  \frac{(\mu^{*})^{-1/2}}{S(\mu^{*} - 1)} - 1  \Big]^{-1}
\end{equation}

Written as above, one sees that the effective mass changes sign when the relevant $f(\mu^{*})$ function equals $\Sigma_{i}\alpha_{i}$. In the isotropic case, $\mu^{*}$=1,
$f(1)=6$, which recovers the $\alpha=6$ limit for the perturbation theory breakdown.
Fig. \ref{fig:alpha_breakdown_uniaxial} shows that the mass breakdown limit is reached earlier when moving away from the isotropic case ($\mu^{*}=1$). In the low physical limit of $\mu^{*}$, the breakdown is dominated by the mass enhancement to the in-plane electronic effective mass, $m^{*}_{\perp}$, reaching a limit as low as 4, while in the high physical limit the limitation comes from the out-of-plane mass enhancement, with an even lower limit of 2. We recall that, as in the standard \fro model, the breakdown of perturbation theory does not necessarily entail a true divergence or pathology of the physical system.

\subsection{Effective masses in the three-fold degenerate case}
\label{sec:deg_eff_mass}

The top valence bands of many cubic materials presents a three-fold degeneracy.
The degeneracy will be lifted when the wavevector departs
from the band extremum (splitting either 2+1 or 1+1+1 depending on the direction). This behaviour follows the general
description of band structures
around degenerate extrema,
given by Luttinger and Kohn~\cite{Luttinger1955}.
Taking the extremum eigenvalue as a reference,
the bare electronic energy dispersion is described by
\begin{equation}
    E_{n}(\Vec{k})= \sigma\frac{k^{2}}{2m_{n}^*(\Vec{k})},    
\end{equation}
up to quadratic order, with a direction- and band-dependent effective mass $m_{n}(\hat{k})$. Such an effective mass fulfills the following eigenvalue equation,
\begin{equation}
    H_{LK}(\Vec{k}) \Vec{e}_{n}(\hat{k}) = \frac{k^{2}}{2m_{n}^*(\hat{k})}\Vec{e}_{n}(\hat{k}),
\label{eqn:eigen_eq_HLK}
\end{equation}
where $H_{LK}$ is the Luttinger-Kohn Hamiltonian matrix whose dimension is equal to the degeneracy of the problem, 
$\Vec{k}=k.\hat{k}$ and $n$ is the band index.
This matrix is written referring to a fixed basis set for the wavefunctions 
which forms an irreducible representation of the
symmetry group of the extremum wavevector. 
The actual $\hat{k}$-dependent eigenfunction for band $n$
in direction $\hat{k}$ is obtained from the eigenvector
$\Vec{e}_{n}(\hat{k})$.

Specifically, in the cubic three-band degenerate case,
the Luttinger-Kohn Hamiltonian
writes~\cite{Luttinger1955}
\begin{widetext}
\begin{equation}
    \begin{split}
       H_{LK} (\Vec{k} ) = 
      \begin{pmatrix}
            Ak_{x}^2 + B(k_{y}^2 + k_{z}^2 )    &                       Ck_{x}k_{y} &                       Ck_{x}k_{z} \\
            Ck_{x}k_{y}                         & Ak_{y}^2 + B(k_{z}^2 + k_{x}^2 )  &                       Ck_{y}k_{z} \\
            Ck_{x}k_{z}                         & Ck_{y}k_{z}                       & Ak_{z}^2 + B(k_{x}^2 + k_{y}^2 )  \\
        \end{pmatrix},
    \end{split}
\label{eq:lhh}
\end{equation}
\end{widetext}
with three parameters $A$, $B$, and $C$.

Turning on the electron-phonon coupling, the polaron dispersion relation in the degenerate band case will have the same behaviour, including the same symmetry characteristics, which gives, similarly to Eqs.~(\ref{eqn:E_P_k}) and~(\ref{eqn:E_P_k_effmass}):

\begin{equation}
\begin{split}
    E_{P,n}(\Vec{k}) &  = \sigma\frac{k^{2}}{2m_{n}^*(\Vec{k})} + \Sigma_{n}(\Vec{k}) \\
                     & \approx \Sigma_{n}(\Vec{k} = 0) + \sigma\frac{k^2}{2m_{P,n}^*(\hat{k})} 
\end{split}     
\end{equation}
with
\begin{equation}
    \frac{1}{m^{*}_{P,n}(\hat{k})} = \frac{1}{m^{*}_n(\hat{k})} + \sigma\frac{ d^{2} \Sigma_n(\Vec{k},E_n(\Vec{k}) )}{ dk^{2}} \Big|_{k=0,\Vec{k}=k.\hat{k}}.
    \label{eqn:m_P_deg}
\end{equation}
Due to the electron-phonon interaction, the bare effective electronic mass $m_{n}^*(\Vec{k})$ is replaced by the polaron effective mass $m_{P,n}^*(\Vec{k})$.

The behaviour of the effective masses with respect 
to the direction $\Vec{k}$ is far from trivial, and generates a so-called ``warping" of the band structure, which has been analyzed in detail by Mecholsky and co-workers~\cite{Mecholsky2014}.
For further reference, we list the inverse of effective masses and degeneracies along the 
$(100)$, $(110)$ and $(111)$
high-symmetry directions.

\begin{equation}
\begin{split}
    m^{*-1}_{n}(100) =& 2A \, \, \text{or} \, \, 2B \, \text{(2-fold)}
    \\
    m^{*-1}_{n}(110) =& (A+B+C), (A+B-C) \, \text{or} \, \, 2B
\\
    m^{*-1}_{n}(111) =& \frac{2}{3}(A+2B+2C) \, \, \text{or} 
    \\
    & \frac{2}{3}(A+2B-C) \, \text{(2-fold)}
\end{split}
\label{eqn:effmasses_3dirs}
\end{equation}

The maximal and minimal effective masses are found along these three directions, depending on the
ratio between Luttinger-Kohn parameters.
The self-energy, in the on-mass-shell approach, is similar to that for the non-degenerate case, although one has to generalize the denominator of Eq.~(\ref{eqn:Sigma_z_k}) from scalar quantities to matrix quantities, as follows:

\begin{widetext}
\begin{equation}
\Sigma_n(\Vec{k},E_n(\Vec{k})) = 
\frac{1}{\pi} \sum_{j}
\int d^{3}q \frac{\omega_{jLO}}{4\pi q^{2}\epsilon^{*}_{j}} 
\Vec{e}_{n}^{\phantom{.}T}(\hat{k}) 
\Bigg( 
\frac{k^{2}}{2m_{n}(\Vec{k})}\mathbb{I}
- H(\Vec{k}+\Vec{q})
- \sigma\omega_{jLO} \mathbb{I}
\Bigg)^{-1}
\Vec{e}_{n}(\hat{k}).
    \label{eqn:Sigma_k_Ek_3}
\end{equation}
\end{widetext}


In order to characterize the polaron effective masses, instead of working with a particular band, it is easier to work with the direct generalization of the Luttinger-Kohn Hamiltonian (which is obtained from generic symmetry considerations) to the electron-phonon case, with modified values of the $A$, $B$ and $C$ parameters, namely $A_P$, $B_P$ and $C_P$.

As an important property of the LK Hamiltonian, the sum of the effective mass inverses for the three bands gives the trace of the LK matrix, which amounts to a constant value, namely $(A+2B)k^2$, independently of the wavevector direction. 
This is indeed fulfilled in Eq.~(\ref{eqn:effmasses_3dirs}).
As a consequence the sum of the polaron effective mass inverses also amounts to a constant value, $2(A_P+2B_P)$, irrespective of the direction of the polaron energy dispersion. 

We now examine specific cases. When the Luttinger-Kohn parameter $C$ vanishes, the electronic dispersion relation corresponds to the absence of coupling between the top(bottom) 3 valence(conduction) bands: the $H_{LK}$ matrix is diagonal.
The eigenvectors of Eq.~(\ref{eqn:eigen_eq_HLK}) are unit vectors for bands $1$, $2$, and $3$, irrespective of the direction $\hat{k}$. The electronic effective masses for the three disjoint bands are determined by the values of the remaining Luttinger parameters, with $m_{\perp}^{-1} = 2B$ and $m_{z}^{-1} = 2A$,
interestingly falling back on the non-degenerate uniaxial case, for each band. 
Thus the analysis of the previous Sec.~\ref{sec:non_deg_eff_mass} applies.

In contrast, if $A=B+C$, one obtains the isotropic three-fold degenerate case, in which, irrespective of the wavevector direction, there is a non-degenerate band, with inverse effective mass $m_0^{-1}=2A$ and two degenerate bands with effective mass $m_1^{-1}=2B$.
This case has been  tackled by Trebin and R\"ossler,\cite{Trebin1975} who provide the analytical expression for the self-energy and effective masses in the one-phonon branch hypothesis. 
The present formalism delivers exactly the same analytical expressions for the effective masses in such case, but also
generalizes them to the multi-phonon case.
Explicitly,
\begin{equation}
    \text{ZPR} = - \frac{\sigma}{3} ((m_0^*)^{1/2}+2(m_1^*)^{1/2}) V_1, 
\label{eqn:zpr_Trebin}
\end{equation} in line with 
Eq.~(\ref{eqn:zpr_short}), and for the polaron effective masses,
\begin{equation}
    \frac{1}{m_{P,0}^*} = 
    \frac{1}{m_{0}^*}
    +
    \Bigg[\frac{3}{10}(m_0^{*})^{-1/2}
          -\frac{2}{15}(m_1^{*})^{-1/2}
          -\frac{(m_1^*)^{1/2}}{3m_0^*}
    \Bigg] 
    V_0, 
\label{eqn:mP0_Trebin}
\end{equation}
and
\begin{equation}
    \frac{1}{m_{P,1}^*} = 
    \frac{1}{m_{1}^*}
    +
    \Bigg[-\frac{1}{15}(m_0^{*})^{-1/2}
          +\frac{1}{15}(m_1^{*})^{-1/2}
          -\frac{(m_0^*)^{1/2}}{6m_1^*}
    \Bigg] 
    V_0, 
\label{eqn:mP1_Trebin}
\end{equation}

In all cases except $C=0$ and $A=B+C$, as soon as $n>1$
Eq.~(\ref{eqn:Sigma_k_Ek_3}) must be evaluated numerically, 
first treating the computation of the self-energy as a function of the wavevector, and then taking its second-order derivative, as needed in Eq.~(\ref{eqn:m_P_deg}). 

In order to calculate the self-energy in the degenerate band case, we perform a numerical integration over $\Vec{q}$ expressed in spherical coordinates:
\begin{widetext}
\begin{equation}
\Sigma_n(\Vec{k},E_n(\Vec{k})) = 
\frac{1}{\pi} \sum_{j}
\frac{\omega_{jLO}}{\epsilon^{*}_{j}}
\int_0^{\infty}dq
\frac{1}{4\pi q^{2}}
\int_{4\pi}d\hat{q}
\,
\Vec{e}_{n}^{\phantom{.}T}(\hat{k}) 
\Bigg( 
\frac{k^{2}}{2m_{n}(\Vec{k})}\mathbb{I}
- H(\Vec{k}+\Vec{q})
- \sigma\omega_{jLO} \mathbb{I}
\Bigg)^{-1}
\Vec{e}_{n}(\hat{k}).
    \label{eqn:Sigma_z_k_expanded}
\end{equation}
\end{widetext}
with 
\begin{equation}
    \Vec{q} = q \hat{q}= q 
    \begin{pmatrix}
     \sin{\theta}\cos{\phi} \\
     \sin{\theta}\sin{\phi} \\
     \cos{\theta}
    \end{pmatrix}.
\end{equation}

Regarding the integral over $q \in [0,\infty)$, one could use a homogeneous grid integration method, 
with the maximal value 
$q_{max}$ tending to infinity. 
However, such homogeneous grid integration approach converges slowly. 
Instead, the semi-infinite domain $q \in [0,\infty)$ can be mapped onto a finite one given that the behaviour of the integrand follows $f(q) \rightarrow \frac{1}{\alpha^{2} + \beta^{2}q^{2} } $, for large $q$. 
One performs the following change of variable
\begin{equation}
    q = \Big( \frac{\omega_{LO}}{\gamma}\Big)^{1/2} \tan{\xi} 
\end{equation}
giving
\begin{equation}
\frac{dq}{d\xi} = \Big( \frac{\omega_{LO}}{\gamma} \Big)^{1/2} \frac{1}{\cos^{2}{\xi}}
=\Big( \frac{\omega_{LO}}{\gamma} \Big)^{1/2}
\Big(\frac{\gamma q^2}{\omega_{LO}}+1\Big)
\end{equation}
and
\begin{equation}
    \int_{0}^{\infty} dq f(q) = \int_{0}^{\pi/2} d\xi \frac{dq}{d\xi} f(q(\xi)),
\end{equation}
\noindent
which can easily be integrated numerically, as the
additional factor in the integrand, $\frac{dq}{d\xi}$,
counterbalances the behaviour of $f(q)$, giving a smooth integrand in the interval $[0,\pi/2]$.
$\gamma$ can be chosen arbitrarily, but the integral converges 
quickly if
it is chosen as the $A$ Luttinger parameter.

For the $\phi$ integration, we use $N_{\phi}$ points,
regularly spaced, with appropriate weights,
while the integral over $\theta$ is performed in terms of a uniform grid of $\cos(\theta)$ using $N_{\theta}$ points, also with appropriate weights.
The convergence values for each studied material are provided in the supplementary material Table S4.







\section{Polaron localization lengths in cubic materials}
\label{sec:localization}

Similarly to the case of effective masses, we will examine
the non-degenerate situation first, then the degenerate case, with particular focus on the three-fold degeneracy.

\subsection{Polaron localization lengths in the non-degenerate case}
\label{sec:localization_non-deg}

In Sec.~\ref{sec:theory_Fr} the polaron radius for the isotropic electronic dispersion was considered, for the case with only one phonon branch. In the present section, the electronic dispersion
is anisotropic, with three different effective masses along the three principal axes, and several phonon branches are included.

The variational energy is again a sum of electronic and $epp$ contributions, as written in Eq.~(\ref{eqn:E_P_var}).
Due to the anisotropy of the effective masses, $E_{el}$
becomes
\begin{equation}
E_{el}[\phi(\Vec{r})]=\int d\Vec{r}
\sum_i
\frac{1}{2m_i} \Big| \frac{\partial\phi(\Vec{r})}{\partial r_i} \Big|^2
\label{eqn:E_el_aniso}
\end{equation}
for $i=x,y,z$.
The density construction,
Eqs.~(\ref{eqn:rho_r}-\ref{eqn:rho_k}), is unchanged.
The $epp$ energy Eq.~(\ref{eqn:E_epp_iso}) now includes 
contributions from several phonon modes, labeled $j$:
\begin{equation}
E_{epp}[\phi(\Vec{r})]=-\sum_j \int d\Vec{k}
\frac{4 \pi}{2k^2} (\epsilon_j^*)^{-1}
\rho^*(\Vec{k}) \rho(\Vec{k})
.
\label{eqn:E_epp_multi}
\end{equation}
The multiphonon situation is actually quite easy to treat in this strong-coupling approach, as one can define 
\begin{equation}
(\epsilon^*)^{-1}=\sum_j (\epsilon_j^*)^{-1},
\label{eqn:multiphonon_epsilon}
\end{equation}
and recover Eq.~(\ref{eqn:E_epp_iso}).

Following the same line of thought as in Sec.~\ref{sec:theory_Fr},
a (anisotropic) Gaussian variational wavefunction ansatz can be considered as follows:
\begin{equation}
    \phi(x,y,z) = \Bigg( \frac{1}{\pi^{3/2}a_{P}^3} \Bigg)^{1/2}
    e^{-\frac{1}{2}
    \Big(\frac{x^{2}}{a_{Px}^2}
         +\frac{y^{2}}{a_{Py}^2}
         +\frac{z^{2}}{a_{Pz}^2}
    \Big)
 }
  ,
\end{equation}
where $a_{Px}$, $a_{Py}$ and $a_{Pz}$ are characteristic localization lengths
along the $x$, $y$ and $z$ directions,
and $a_{P}$ is the volumetric 
mean radius of the polaron, that is, the radius of the sphere that gives the same volume as that of the ellipsoid built on the semi-axes $a_{Px}$, $a_{Px}$ and $a_{Px}$,
\begin{equation}
    a_{P}^3=a_{Px}a_{Py}a_{Pz}.
    \label{eqn:a_PV}
\end{equation}
With such a trial wavefunction, $E_{el}$ and $E_{epp}$ become
\begin{equation}
E_{el}=\frac{1}{4}    
    \Big(\frac{1}{a_{Px}^2 m_x^*}
        +\frac{1}{a_{Py}^2 m_y^*}
        +\frac{1}{a_{Pz}^2 m_z^*}
    \Big)
\label{eqn:E_el_aniso_gaussian}
\end{equation}
and
\begin{eqnarray}
E_{epp}&=&-\frac{1}{\epsilon^* (2\pi)^{1/2}} 
    \nonumber \\
&&
    \times \frac{1}{4\pi}
    \int_{4\pi}d\hat{k}
    \Big( a_{Px}^2\hat{k}_x^2
         +a_{Py}^2\hat{k}_y^2
         +a_{Pz}^2\hat{k}_z^2
    \Big)^{-1/2}.
    \nonumber \\
\label{eqn:E_epp_aniso_gaussian}
\end{eqnarray}
A polaron directional localization length $a_{Pd}(\hat{k})$ is
then defined as 
\begin{equation}
a_{Pd}(\hat{k})= \Big(
          a_{Px}^2\hat{k}_x^2
         +a_{Py}^2\hat{k}_y^2
         +a_{Pz}^2\hat{k}_z^2
            \Big)^{1/2},
\label{eqn:a_P_hat_k}
\end{equation}
the angular average of its inverse being
\begin{equation}
<a_{Pd}^{-1}>_{4\pi}= \frac{1}{4\pi}
\int_{4\pi} d^3k \, (a_{Pd}(\hat{k}))^{-1},
\label{eqn:a_P_m1_avg}
\end{equation}
so that $E_{epp}$ can be written in terms of this average,
\begin{eqnarray}
E_{epp}&=&-\frac{<a_{Pd}^{-1}>_{4\pi}}{\epsilon^* (2\pi)^{1/2}} .
\label{eqn:E_epp_aniso_short}
\end{eqnarray}

Minimization of $E_{el}+E_{epp}$ by variation of the localization lengths $a_{Pi}$, for given $\epsilon^*$ and $m_i^*$ delivers the optimal Gaussian trial wavefunction.

This optimization problem does not have an analytical solution in the most general case, but can be treated numerically, either by direct minimization of $E$ in the space of $a_{Pi}$, or by requiring that the derivatives of $E$ with respect to $a_{Pi}$ vanish, which amounts to 
\begin{equation}
    \frac{1}{m^*_i}=-\frac{2a_{Pi}^3}{\epsilon^*(2\pi)^{1/2}}
    \frac{\partial <a_{Pd}^{-1}>_{4\pi}}{\partial a_{Pi}}.
\end{equation}

As for the effective masses, a specific treatment of the uniaxial case is possible. It does not deliver
explicit values for the localization lengths as a function of $\epsilon$ and $m_i^*$, but an implicit relationship, as follows. 
The convention for the $x$ and $y$ masses are given by Eq.~(\ref{eqn:m_uniaxial}). 
By symmetry, the $x$ and $y$ localization lengths must be equal, with 
\begin{equation}
    a_{Px} = a_{Py} = a_{P\perp},
    \label{eqn:a_uniaxial}
\end{equation}
We define the anisotropy parameter $\delta$:
\begin{equation}
\delta =\frac{a^{2}_{Pz}}{a^{2}_{P\perp}} - 1,
\label{eqn:delta}
\end{equation}
and arrive at relatively simple and intuitive formulas for the in-plane, $a_{\perp}$, respectively out-of-plane, $a_{z}$, polaron localization lengths as functions of 
$a_{P}$ and $\delta$:

\begin{equation}
    \label{eqn:apperp}
    a_{P\perp} = a_{P} ( 1 + \delta )^{-1/6}
\end{equation}

\begin{equation}
    \label{eqn:apz}
    a_{Pz} = a_{P} ( 1 + \delta )^{1/3}
\end{equation}
After some algebra, the energy to be minimized as a function of $a_{P}$ and $\delta$ writes

\begin{eqnarray}
E_{P}&=&\frac{1}{4a_{P}^2}    
    \Big(\frac{2( 1 + \delta )^{1/3}}{m_\perp^*}
        +\frac{( 1 + \delta )^{-2/3}}{m_z^*}
    \Big)
    \nonumber\\
&-&\frac{1}{\epsilon^* (2\pi)^{1/2}a_{P}} ( 1 + \delta )^{1/6} S(\delta),
\label{eqn:E_P_aniso_intermediate}
\end{eqnarray}
where the $S$ function is defined in Eq.~(\ref{eqn:sfunction}).

Optimization with respect to $a_{P}$ is easy, while 
the determination of $\delta$ is more delicate. Considering $\delta$ as the basic variable, the effective mass ratio $\mu^{*} = \frac{m_{\perp}}{m_{z}}$ can be determined analytically:
\begin{equation}
    \mu^{*}(\delta) = 
    S \Big(\frac{dS}{d\delta}\Big)^{-1}  + \frac{1}{2(1+\delta)} - 2(1+\delta),
    \label{eqn:mustar}
\end{equation}
where the derivative of the $S$ function is
\begin{equation}
    \label{eqn:dsfunction}
    \frac{dS}{d\delta} = \frac{1}{2\delta(1+\delta)^{1/2}} - \frac{S(\delta)}{2\delta}.
\end{equation}
An explicit solution
of the optimization problem would require $\delta$ as a function of $\mu^*$, but unfortunately Eq.~(\ref{eqn:mustar}) cannot be inverted analytically.
Still, the polaron localization length and energy
can be determined as a function of $\delta$ as well,
so that there is a parametric relationship between
the polaron characteristics and $\mu^{*}$.

Compared to the spherical (isotropic) scenario, 
Eq.~(\ref{eqn:ap_sphere}), the spherical average of the polaron localization length is now:
\begin{equation}
    a_{P} = \frac{3}{2} \sqrt{2\pi} <m^{*1/2}>^{-2} \epsilon^{*} D(\delta)
\label{eqn:ap_spherical}    
\end{equation}
\newline
where $D(\delta)$ is the enhancement factor with respect to the isotropic case. Its expression is

\begin{equation}
    \label{eqn:dfactor}
    D(\delta) = \frac{1}{3} (1 + \delta)^{-5/6} 
    \Bigg( 
    \frac{dS}{d\delta} + \frac{S(\delta)}{2(1+\delta)}  
    \Bigg)^{-1}.
\end{equation}

\begin{figure}
    \centering
    \includegraphics[width=0.48\textwidth]{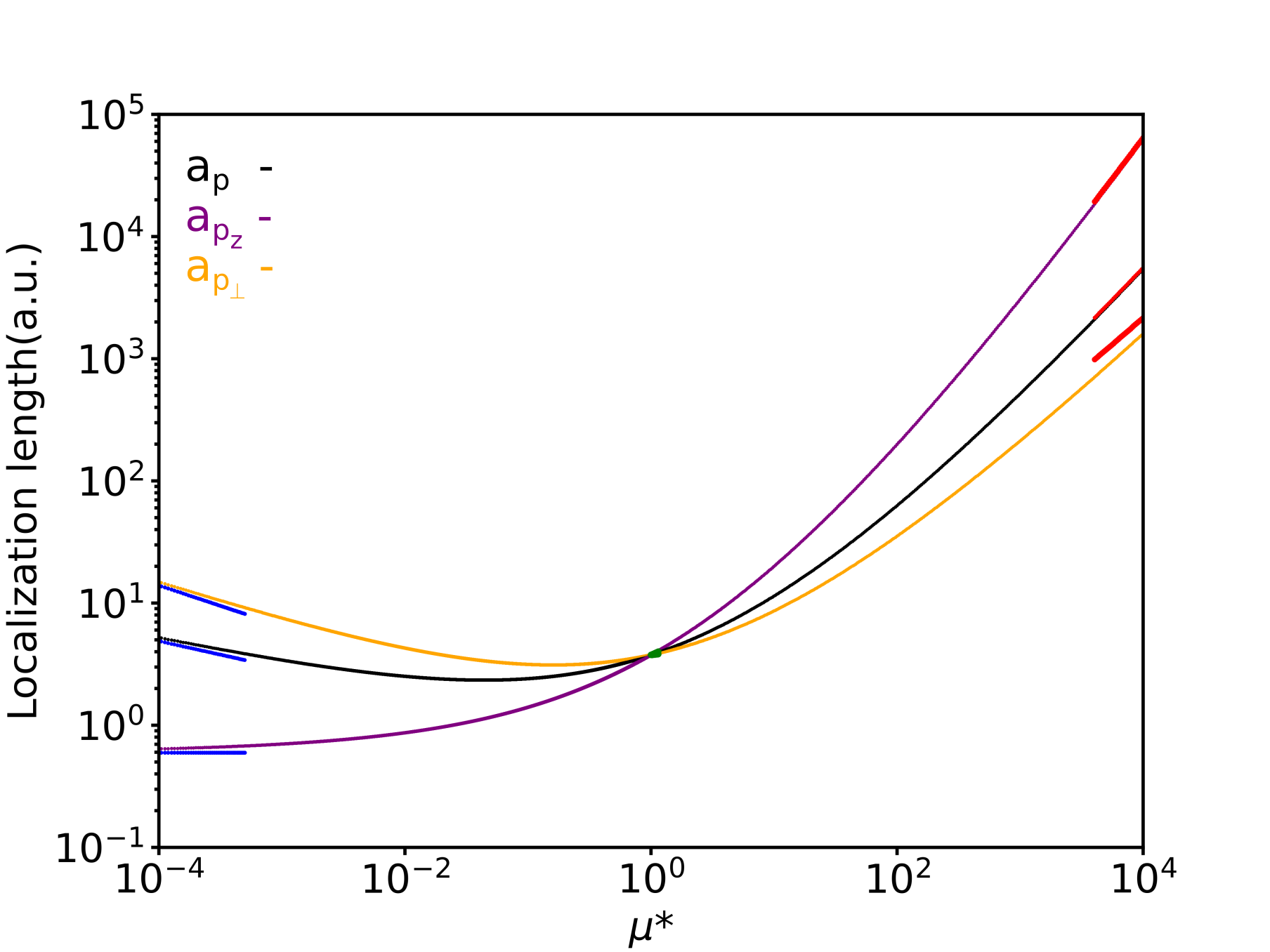}
    \caption{Polaron localization length in the uniaxial limit calculated at constant density of states expressed in terms of $\mu^{*}$ for the volumetric average case ($a_{P}$), the out-of-plane localization length ($a_{Pz}$), respectively the in-plane localization length ($a_{P\perp}$). In the low limit of $\mu^{*}$ , the asymptotic analytical behaviour is represented in blue. In the $\mu^{*}\rightarrow 1$ limit, the analytical expression is represented in green. Finally, in the large $\mu^{*}$ limit, the asymptotic behaviour analytical limit is represented in red. The expressions for the analytical asymptotic behavior in the three cases are provided in the supplementary material. We consider here $\epsilon^{*} = 1$.}
    \label{fig:ap_vs_mu_uniaxial}
\end{figure}




In Fig. \ref{fig:ap_vs_mu_uniaxial} we show the in-plane, $a_{p_{\perp}}$, respectively out-of-plane, $a_{p_z}$, localization lengths expressed in terms of $\mu^{*}$. For $\mu^{*}$ smaller than 1, the smallest polaron localization length is $a_{p_z}$, while for $\mu^{*}$ larger than 1, $a_{p_\perp}$ is the smallest. For $\mu^{*} = 10^{4}$, the value of $a_{p_\perp}$ is roughly one order of magnitude smaller than $a_{p_z}$.

Finally the polaron formation energy in the uniaxial case can be expressed in terms of $\delta$ and $\mu^*$:

\begin{equation}
    E(\delta) = -4\pi^{3} \epsilon^{*-2} m^{*}_{\perp} (S(\delta))^{2} \Big( 1 + \frac{\mu^{*}}{2(1+\delta)}  \Big)^{-1}.
\label{eqn:e_uniaxial}
\end{equation}
\noindent

\begin{figure}
    \centering
    \includegraphics[width=0.48\textwidth]{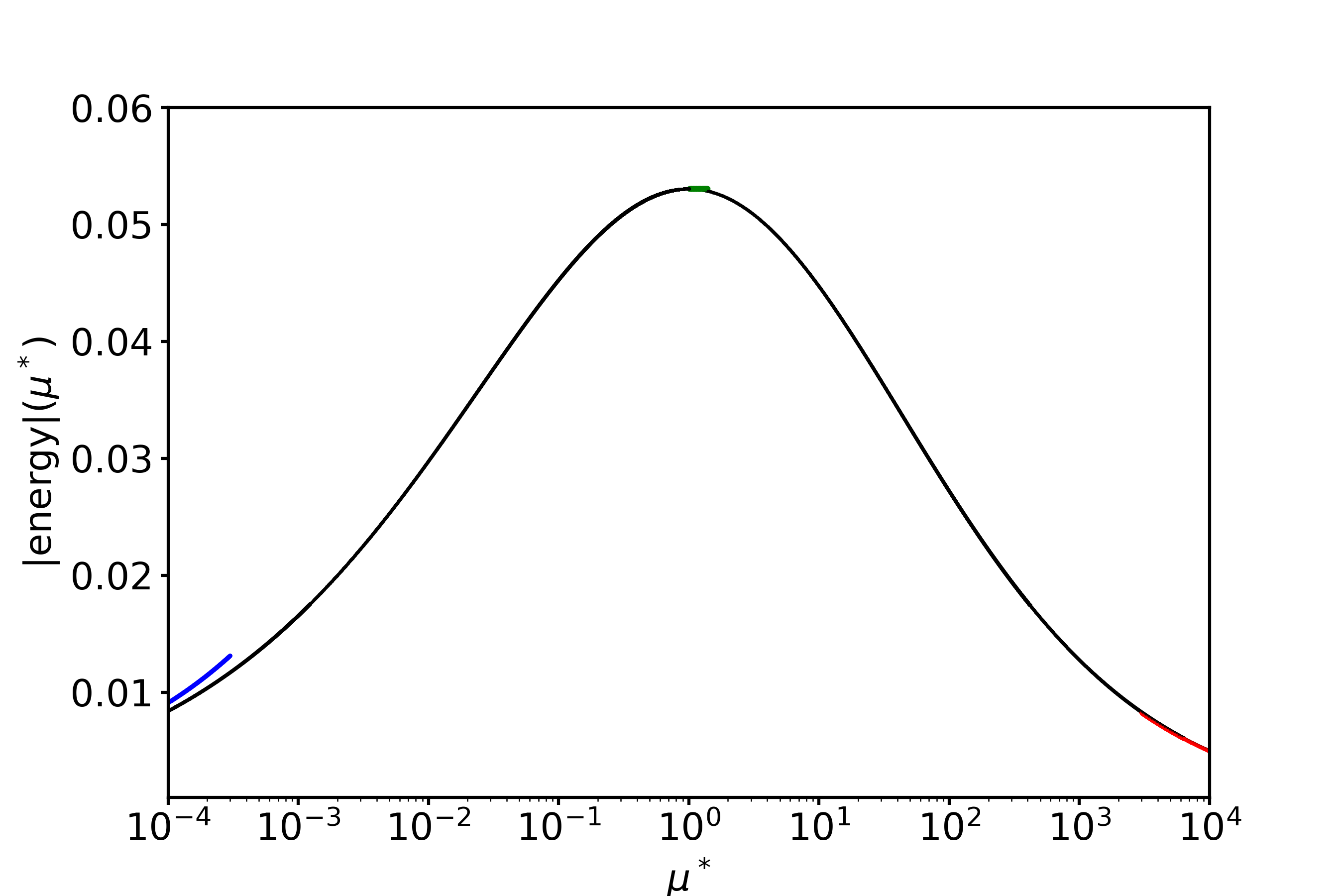}
    \caption{Absolute value of the polaron formation energy in the uniaxial limit calculated at constant volumetric localization length as a function of $\mu^{*}$. In the low limit of $\mu^{*}$ , the asymptotic analytical behaviour is represented in blue. In the spherical limit, the analytical expression is represented in green. Finally, in the large $\mu^{*}$ limit, the asymptotic behaviour analytical limit is represented in red. The expressions for the analytical asymptotic behavior in the three cases are provided in the supplementary material. We consider $\epsilon^{*} = 1$.}
    \label{fig:e_vs_mu_uniaxial}
\end{figure}
Numerically, it is then easy to obtain $E$ as a function of $\mu^*$ only,
which is represented
in Fig. \ref{fig:e_vs_mu_uniaxial}.

\subsection{Polaron localization lengths in the degenerate case}

In the degenerate case, the polaronic wavefunction includes several components, namely, a superposition
of contributions from the different degenerate electronic bands.
We introduce the band label $j$ that runs from $1$ to $n$, and note the components of the wavefunction $\phi_j(\Vec{r})$. 
The electronic density is formed by summing the densities from all components,
\begin{equation}
\rho(\Vec{r})=\sum_j|\phi_j(\Vec{r}) \big|^2,
\label{eqn:rho_r_deg}
\end{equation}
and contributes to the $E_{epp}$ in the same way as in the non-degenerate case, even if there are several phonon branches.
The electronic energy, $E_{el}$, is also obtained from the contribution of the different components, through the Luttinger-Kohn Hamiltonian.
Expressed in reciprocal space,
\begin{equation}
E_{el}[\phi_j]=\sigma \int d^3{k}
\sum_{j,j'=1}^{n} \phi_j^*(\Vec{k}) H_{LK}(\Vec{k})\Big|_{(j,j')} \phi_{j'}(\Vec{k}).
\label{eqn:E_el_deg}
\end{equation}
Moreover, the multicomponent wavefunction must be normalized, 
\begin{equation}
1= \sum_{j=1}^{n} \int d^3{k}
 \phi_j^*(\Vec{k}) \phi_j(\Vec{k})=
\sum_{j=1}^{n} \int d^3{r}
 \phi_j^*(\Vec{r}) \phi_j(\Vec{r}).
\label{eqn:norm_multicomponent}
\end{equation}

Like in the non-degenerate case, we are not looking for an exact solution of this minimization problem, but we are trying to find the best trial wavefunction among a set of acceptable wavefunctions with some adjustable parameters. 

In particular, since the $E_{epp}$ contribution
depends only on the density, and not on the way 
this density is generated from the different components, the same $E_{epp}$ is obtained if the
density is generated entirely from one component only, e.g. by hypothesizing that this component of the wavefunction 
is the square root of the density, or whether this component is generated from the same square root 
multiplied by some fraction for one component, some other fraction for a second component, etc, or even by a less regular spread of the density over the components. 
However, one can hypothesize that in order to minimize the kinetic energy (or equivalently, 
as a consequence of the quadratic behaviour of the $H_{LK}$ Hamiltonian with respect to the wavevector length), the $E_{el}$ will be minimal when all the components
have the same spatial variations. 
Hence, one is lead to the following ansatz,
\begin{equation}
\phi_j(\Vec{r})=c_j \phi_0(\Vec{r}),
\end{equation}
where the coefficient $c_j$ are normalized,
\begin{equation}
1=\sum_j |c_j|^2
\end{equation}
as well as the wavefunction $\phi_0(\Vec{r})$:
\begin{equation}
1= \int d^3{k}
 \phi_0^*(\Vec{k}) \phi_0(\Vec{k})=
\int d^3{r}
 \phi_0^*(\Vec{r}) \phi_0(\Vec{r}).
\label{eqn:norm_multicomponent0}
\end{equation}
The density is written directly from $\phi_0(\Vec{r})$,
\begin{equation}
\rho(\Vec{r})=|\phi_0(\Vec{r}) \big|^2.
\label{eqn:rho_r_deg_0}
\end{equation}

In this case, the electronic energy becomes
\begin{equation}
E_{el}[\phi_0,c]=\sigma \int d^3{k}
\phi_0^*(\Vec{k}) \phi_{0}(\Vec{k})
\sum_{j,j'=1}^{n} c_j^* H_{LK}(\Vec{k})\Big|_{(j,j')} c_j.
\label{eqn:E_el_deg_split}
\end{equation}
Morover, $\sum_{j,j'=1}^{n} c_j^* H_{LK}(\Vec{k})\Big|_{(j,j')} c_j$ is a quadratic
form over the wavevectors, that has principal axes. Along these principal axes one recovers
the usual dispersion relation
\begin{equation}
    E_{kin}(\Vec{k})= 
    \frac{k_1^{2}}{2m_1^*}+
    \frac{k_2^{2}}{2m_2^*}+
    \frac{k_3^{2}}{2m_3^*}
\end{equation}
although the new $1, 2$ and $3$ principal axes might not be aligned
with the initially chosen $1, 2$ and $3$ states.

The search for the best trial wavefunction
can proceed as follows:
(1) one chooses several possible $c_j$ sets;
(2) for each of these sets, one solves a non-degenerate optimization problem, of the kind 
already tackled in Sec.~\ref{sec:localization_non-deg}, with effective masses $m_n^*$, for which the polaron formation energy can be computed as well as the localization length;
(3) among the different $c_j$ sets, the best ansatz will be the one that delivers the lowest polaron formation energy, with corresponding localization length.

In practice, we consider for $c_j$, the three directions $(100)$, $(110)$ and $(111)$, for which we know the effective masses, see Eq.~(\ref{eqn:effmasses_3dirs}).
For the $(100)$ and $(111)$ directions, we recover the non-degenerate uniaxial case, as two of these effective masses are equal. However, for the 
$(110)$ direction, the three effective masses
are distinct, and the variational problem must be addressed entirely numerically, or using an approximate expression for $<a_{Pd}^{-1}>_{4\pi}$,
that we provide in the supplementary information.

\section{Results}
\label{sec:result}

We have applied our new formalism to a range of (cubic) materials, including II-VI compounds: CdS, CdSe, CdTe, ZnS, ZnSe, ZnTe; III-V compounds: AlAs, AlSb, AlP, BAs, BN, GaAs, GaN, GaP; oxides: BaO, CaO, Li$_2$O, MgO, SrO; and SiC, adding up to a total of 20 cases. In all the cases we study both hole and electrons polarons whose characteristics are analyzed in the present section.

We find a range of polaron behaviors in the studied materials among which we identified: 9 electron polarons in the non-degenerate and isotropic case, 11 electron polarons and one hole polaron in the non-degenerate and anisotropic case, and 19 hole polarons in the 3-fold degenerate case.

In order to obtain parameters for the gFr model, we use Density Functional Theory (DFT) as implemented in the ABINIT package\cite{Gonze2020abinit, Gonze2016,Gonze1997}.
Calculation parameters (kinetic energy cut-off and reciprocal space sampling) for both the electronic and phononic wave-vectors are provided in Table S1. The exchange-correlation energy has been treated using the GGA-PBE approximation\cite{Perdew1996} with the corresponding norm-conserving pseudo-potentials provided by the PseudoDojo project\cite{Setten2018}, except for BN-zb where previously published results were used\cite{Ponce2015}.
Based on the converged electronic wavefunctions, Density Functional Perturbation Theory (DFPT)\cite{Gonze1997,Laflamme2016} has been used in order to obtain the Luttinger parameters, electronic effective masses, dielectric constants and $\Gamma$-point phonon frequencies. Parameters can be found in Tables  S2 and S3.

\begin{figure}[!hptb]
    \centering
    \includegraphics[width=0.45\textwidth]{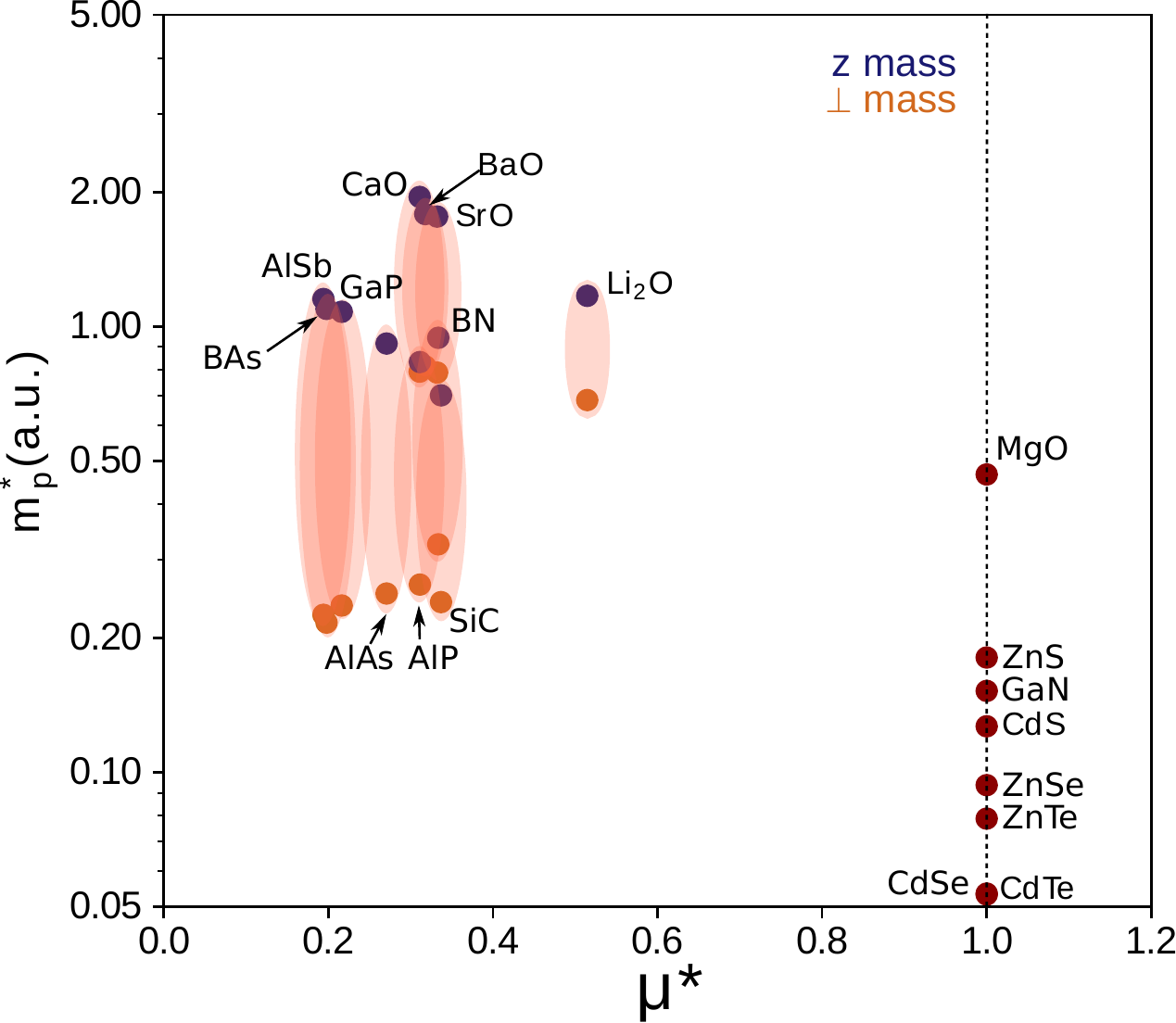}
    \caption{Electron polaron effective mass as a function of the ratio between in-plane, $m_\perp$, and out-of-plane, $m_z$, bare electron effective masses for the studied materials. For the full set of values refer to Table S12.}
    \label{fig:cb_mp_vs_mu}
\end{figure}

\begin{figure}[!hptb]
    \centering
    \includegraphics[width=0.475\textwidth]{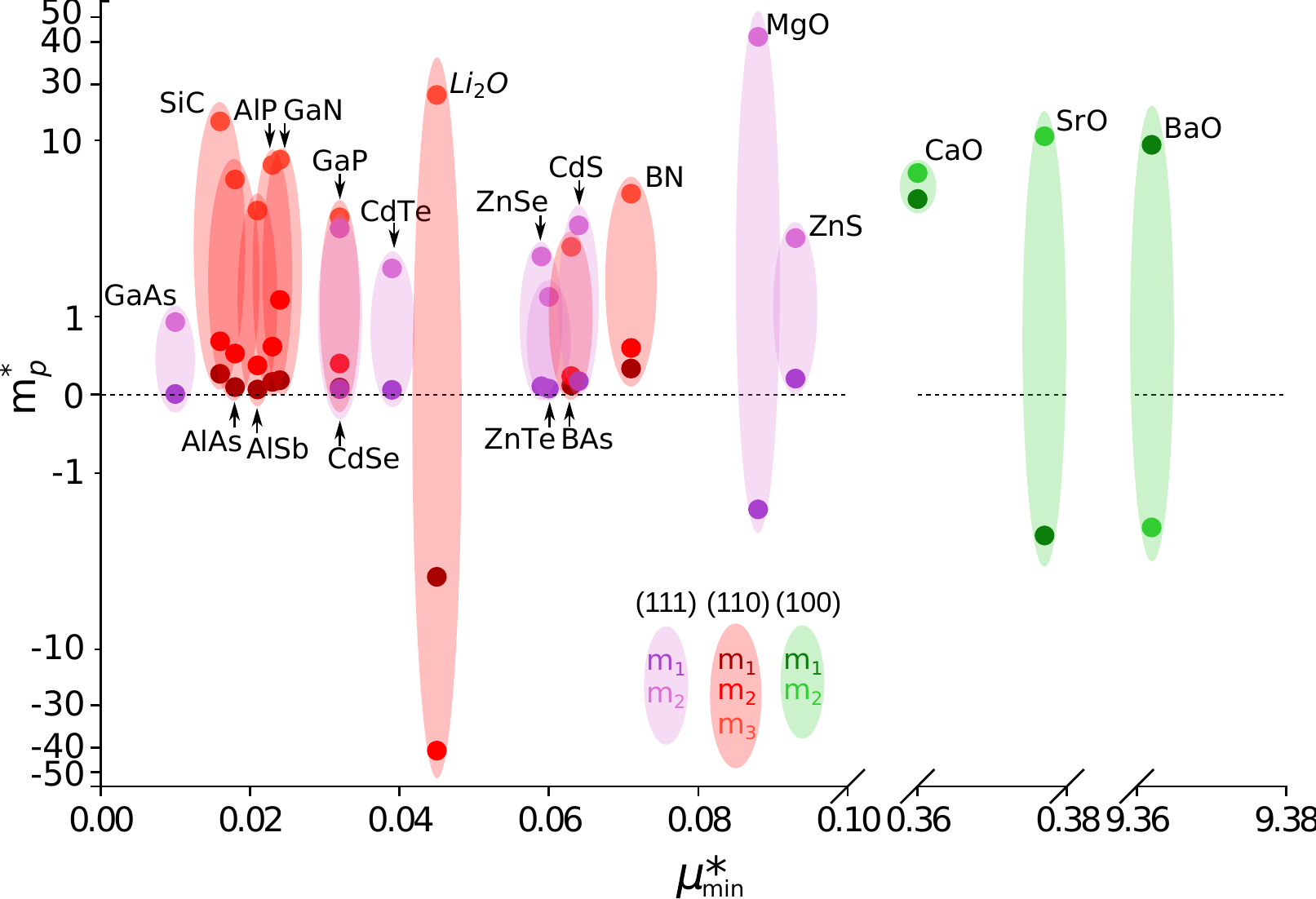}
    \caption{Hole polwaron effective mass as a function of the (minimum) ratio between the corresponding electronic effective masses for the studied materials. Negative values for the polaron effective masses indicate a crossing of the mass breakdown limit. For the full set of values refer to Table S6.} 
    \label{fig:vb_mp_vs_mu}
\end{figure}

We show the corresponding electron and hole polaron effective masses in Fig.~\ref{fig:cb_mp_vs_mu} and~\ref{fig:vb_mp_vs_mu}. 
For the non-degenerate case, this is presented as a function of the ratio between the in-plane, $m_{\perp}$, and out-of-plane, $m_{z}$, electronic effective masses, referred to as $\mu^{*}$. 
The reason for which we depart from the spherical shape follows from the character of the electronic band structure. 
While for the spherical effective masses the conduction band minimum is at the Brillouin zone center, in the ellipsoidal case the minimum is found at $X$ or $L$. 
For the degenerate case, we also use an anisotropy measure to plot the results,
namely the minimum ratio between the electronic effective masses of different bands, 
be it along the (100), the (110) or the (111) direction, depending on the direction along which the polaron has the lowest formation energy in the strong-coupling case.
We note this quantity $\mu^{*}_{min}$.

As a sanity check of the obtained values, for the degenerate case, we compute the sum of the 3 inverse polaron effective masses
for different directions, and find that it is indeed constant, and independent of the direction in the Brillouin zone. 
This is expected based on the derivations above, from a similar property described in Eq. \ref{eqn:effmasses_3dirs}. 

\begin{figure}[!hptb]
    \centering
    \includegraphics[width=0.425\textwidth]{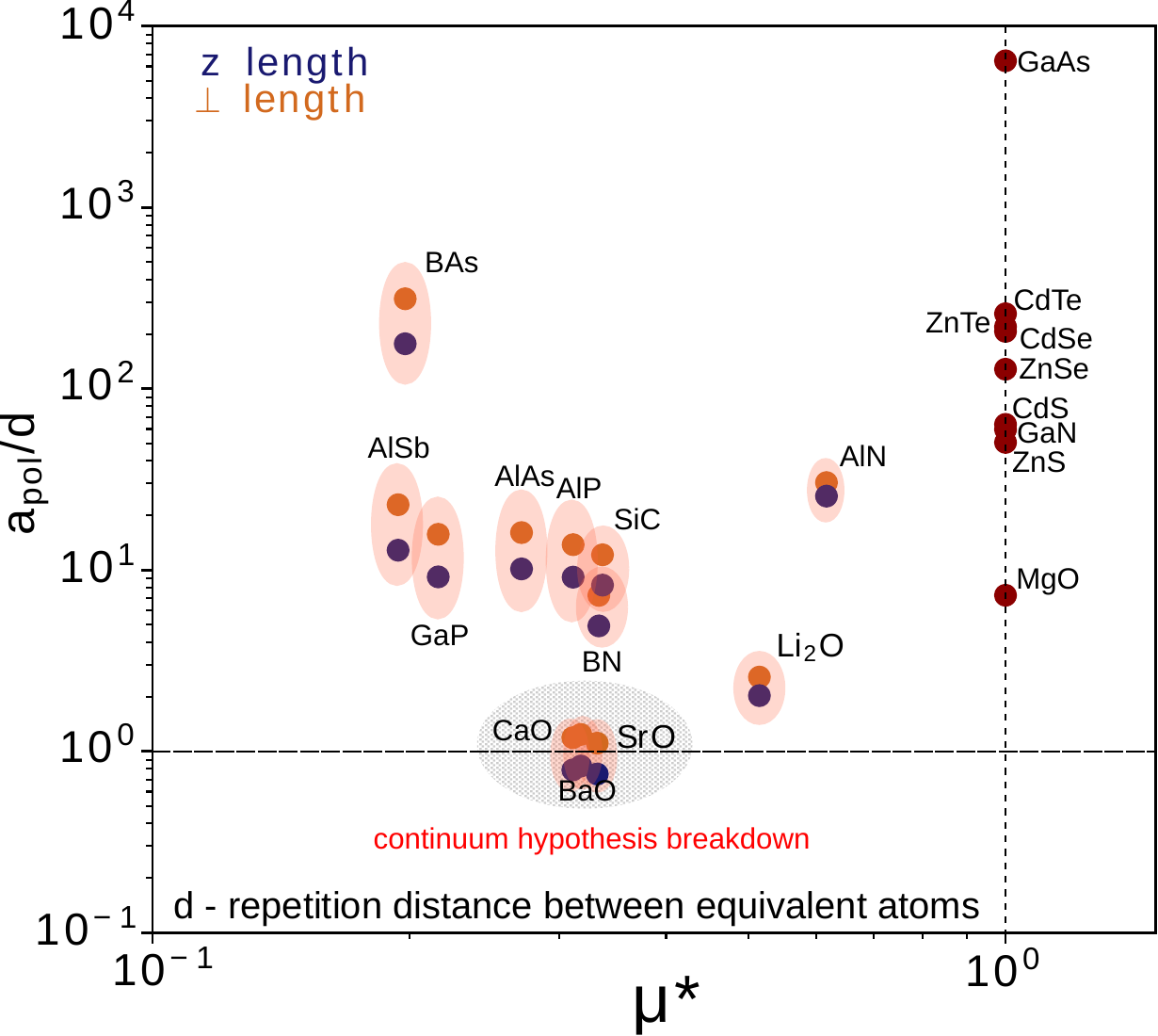}
    \caption{Electron polaron localization length expressed in units of repetition distance $d$ between equivalent atoms as a function of the ratio between in-plane, $m_\perp$, and out-of-plane, $m_z$, effective masses for the studied materials. For the full set of values refer to Table S9.  Note that when the two effective masses are equal, the isotropic case is reached and the characteristic localization length is isotropic (red dots), while for the other cases two localization lengths are necessary, for the in- and out-of-plane lengths(orange and blue dots).
    }
    \label{fig:apc_vs_mustar}
\end{figure}

\begin{figure}[!hptb]
    \centering
    \includegraphics[width=0.45\textwidth]{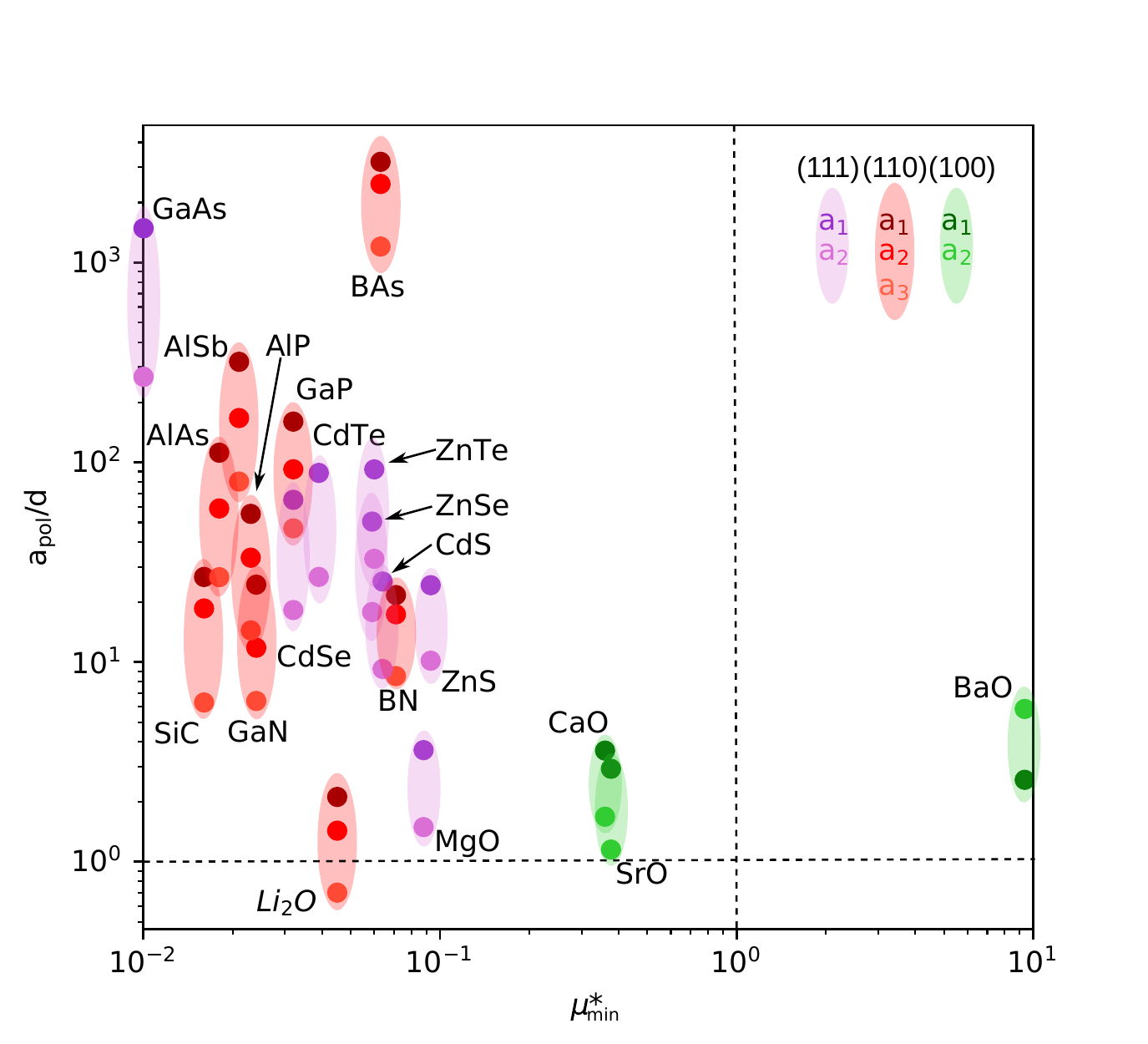}
    \caption{Hole polaron localization lengths along the direction with lowest formation energy, expressed in units of primitive lattice constants as a function of the (lowest) ratio between the bare hole effective masses for the studied materials. For the full set of values refer to Table S8.a. Note that in the case when the lowest polaron formation energy is along the (110) direction there is no degeneracy in the electronic effective masses and, consequently, the effective polaron volume has the shape of a triaxial ellipsoid. $a_{1}$ represents the polaron radius along which the polaron  stabilizes ((100), (110) or (111)), $a_2$ - the polaron radius along the direction perpendicular to the plane containing $a_1$ (i.e. perpendicular to (100), (110) or (111)), while $a_3$ is the radius along the direction perpendicular to the plane containing $a_1$ and $a_2$.}
    \label{fig:hp_apc_vs_mustar_lowest}
\end{figure}

In Fig. \ref{fig:apc_vs_mustar} (resp. \ref{fig:hp_apc_vs_mustar_lowest}) we report the localization length for the electron (resp. hole) polaron, in units of the crystal lattice constant. This is also presented as a function of the ratio between the in-plane, $m_{\perp}$, and out-of-plane, $m_{z}$, electronic effective masses, referred to as $\mu^{*}$. 
If $\mu^{*} = 1$ we recover the spherically shaped polaron scenario, while for other values we describe ellipsoidal shaped polarons with two different axes: $a_{p_{\perp}}$, respectively $a_{p_{z}}$.
More importantly, for some oxides the localization length is on the order of the distance between the atoms, which indicates a breakdown of the \fro continuum hypothesis.

Along the (100) or (111) directions, polarons have a shape of an oblate spheroid for $\mu^{*} < 1$, and a prolate one for $\mu^{*} > 1$. The deviation from a sphere becomes more visible with increasing difference between the 3 radii characterizing the polaron, as we move away from $\mu^{*}=1$. In the case of (110) direction the degeneracy among the effective masses is lifted and the polaron takes the shape of a general ellipsoid.

\section{Discussion}
\label{sec:discussion}

In this section we discuss the
results of the previous section.
First, we examine the possible breakdowns of the basic hypotheses (continuum hypothesis or validity of perturbation theory) of our results, then make some comparisons of our results with those of an approximate 
treatment of anisotropy with the Feynman variational approach, for non-degenerate case.

\begin{figure}[!hptb]
    \centering
    \includegraphics[width=0.475\textwidth]{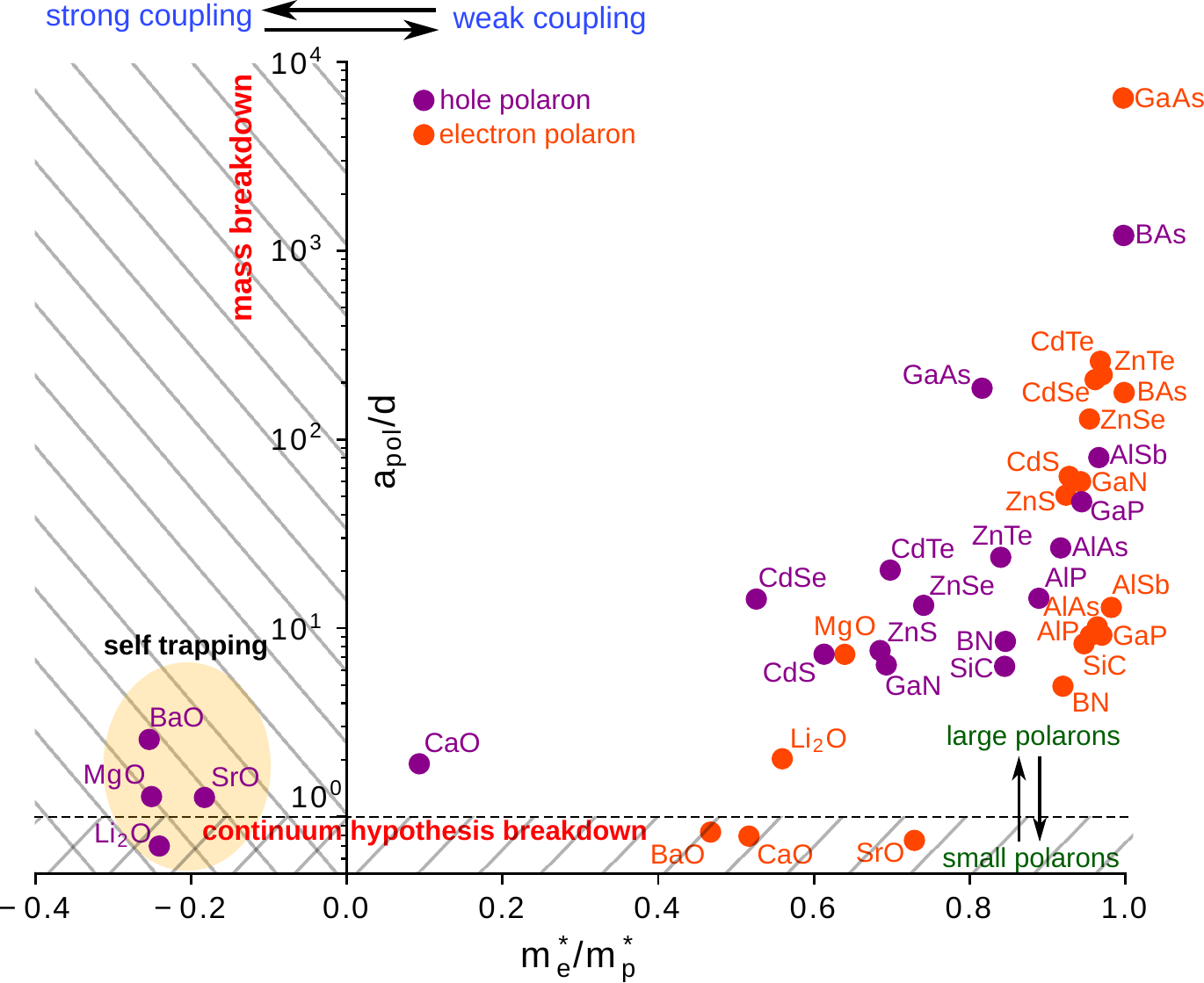}
    \caption{Shortest electron and hole polaron localization lengths, expressed in units of repetition distance between equivalent atoms, as a function of the smallest ratio between carrier (electron/hole) effective mass and polaron effective mass. When the mass ratio drops below 0, perturbation theory breaks down, while when the localization ratio drops below 1, the continuum hypothesis breaks down. For the full set of values refer to Table S8.a and S9.}
    \label{fig:ap_vs_memp}
\end{figure}

We show in Fig. \ref{fig:ap_vs_memp} the characteristics (effective mass and localization length) for both electron and hole polarons in the studied materials.
On the horizontal axis, the ratio between the electronic and polaronic effective masses characterizes the strength of the electron-phonon coupling: the closer the ratio is to $1$, the weaker the coupling, whereas for very strong EPC we arrive at a change of sign for the polaron effective mass. 
As previously discussed this is associated
with a breakdown of weak-coupling perturbative treatment, and indicates that a strong-coupling approach is needed: self-trapped polarons will form in these oxides. 
On the other hand, we set a somewhat arbitrary breakdown threshold (a$_P \approx$ d) to characterize the applicability of the continuum hypothesis inherent to the \fro model (a$_P \gg$ d). 
We show the shortest localization length in cases beyond the non-degenerate isotropic scenario. 

We note that for some of the studied materials the localization length is of the order of the repetition distance (e.g. BaO, SrO, CaO, MgO, Li$_2$O, for both electron and hole polarons, except the MgO electron polaron), and as a consequence we reach the continuum breakdown limit: the discrete character of the crystal should be taken into account, and another theoretical model would be more suitable to describe the polaron.
The weak-coupling perturbation theory treatment 
also breaks down for the hole polarons of 
BaO, SrO, MgO and Li$_2$O. For all these cases,
the ZPR obtained from perturbation theory (including the AHC approach) might be questioned.

The results from the Feynman approach, computed exactly for the isotropic mass situation, are averaged as described in 
Sec.~\ref{sec:Feynman_model}. This is done 
for the non-degenerate anisotropic electronic dispersion.

\begin{figure}
    \centering
    \includegraphics[width=0.475\textwidth]{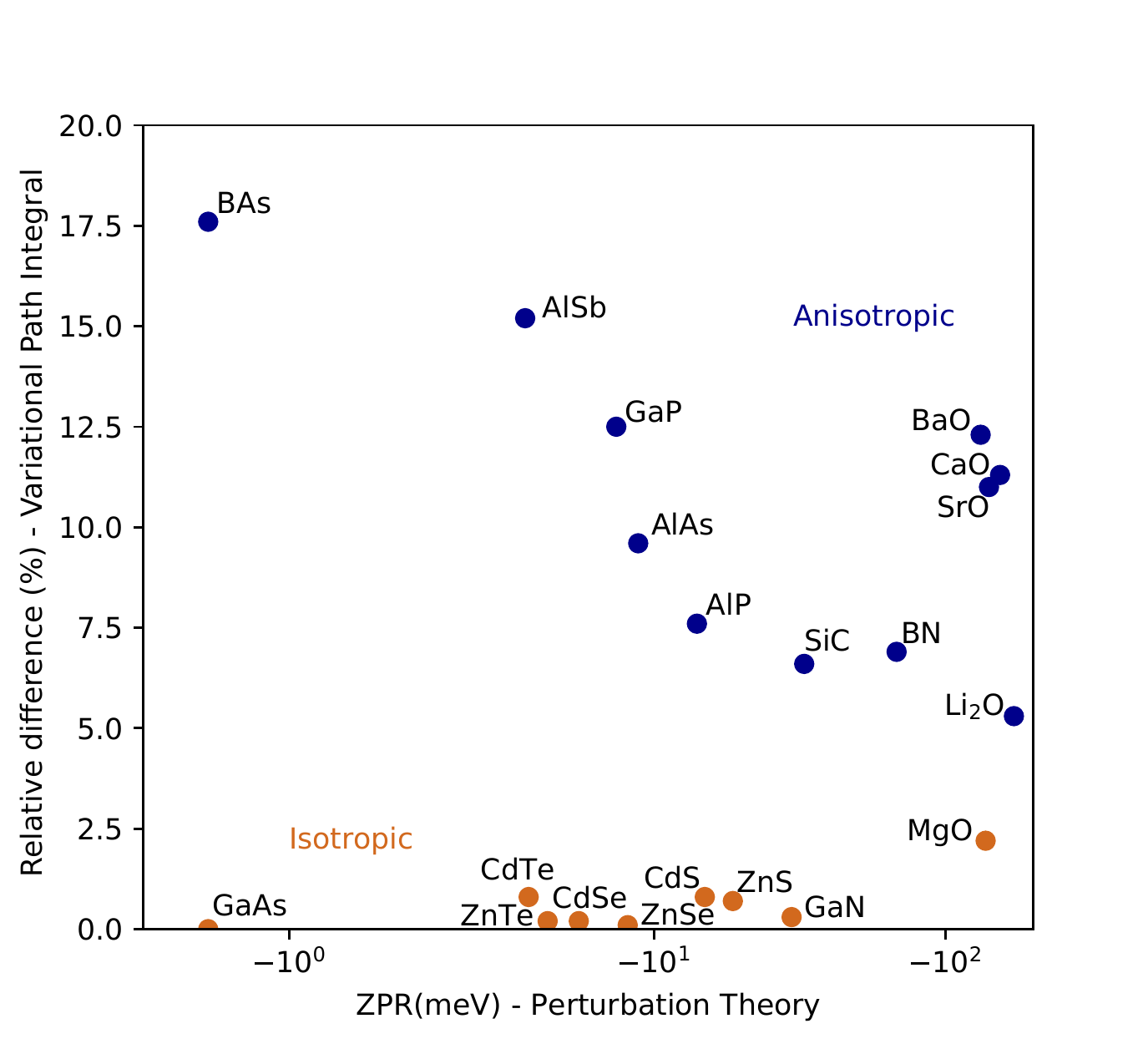}
    \caption{In the non-degenerate electron polaron situation, relative difference between the ZPR determined using perturbation theory (fully taking into account the possible anisotropy), and Feynman variational path integral approach (with approximate treatment of the anisotropy). In all cases the ZPR determined using the perturbative method has a larger value than the ZPR determined using the variational approach. For the full set of values refer to Table S11.}
    \label{fig:ff_zpr}
\end{figure}

In Fig. \ref{fig:ff_zpr} we note that the variational approach reaches, as expected, a lower ground state energy for both anisotropic and isotropic electronic dispersions. 
While in the isotropic case the relative difference between the two theoretical approaches remain within 2.5 $\%$, the anisotropic case produces differences as high as 17.5 $\%$ among the studied materials. Although in both treatments the wavefunction is constructed based on a Gaussian function, in the Feynman approach we consider an approximate treatment of the anisotropy which nevertheless yields a reasonable comparison.

\begin{figure}
    \centering
    \includegraphics[width=0.475\textwidth]{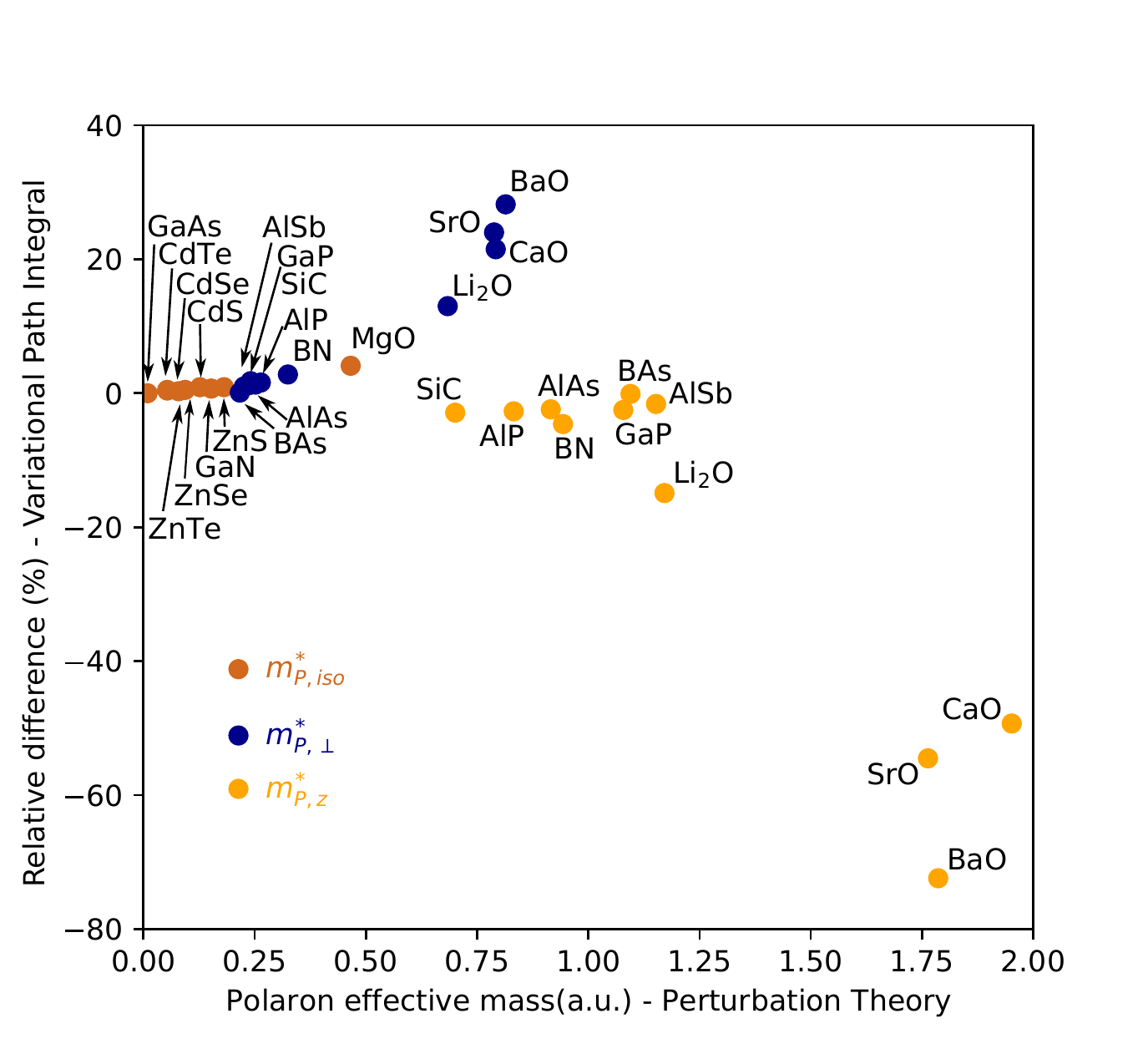}
    \caption{
    In the non-degenerate electron polaron situation, relative difference between the effective masses determined using perturbation theory (fully taking into account the possible anisotropy), and Feynman variational path integral approach (with approximate treatment of the anisotropy). $m^{*}_{P,iso}$ is the isotropic effective mass, while in the anisotropic case $m^{*}_{P,\perp}$ and $m^{*}_{P,z}$ are the in-plane and out-of-plane polaron effective masses.}
    \label{fig:ff_emass}
\end{figure}

In Fig.~\ref{fig:ff_emass} we present the relative difference in polaron effective mass between the above-mentioned theoretical approaches. We note that the largest difference is found in materials that, within the \fro approach, are at the continuum limit breakdown, like BaO, CaO, SrO and, to a lesser extent, Li$_2$O. In both anisostropic and isotropic cases we note that the relative difference increases with polaron effective mass, and the in-plane and out-of-plane effective mass differences seem to  diverge. This sudden increase in the relative difference is associated with the lower $\alpha$ breakdown limit in the anisotropic case, as expressed in Fig. \ref{fig:alpha_breakdown_uniaxial}.

\section{Conclusions}
\label{conclusion}

In this work, we propose a formalism to describe realistic scenarios of polarons in cubic materials. Starting from the basic \fro model, we include both anisotropic and degenerate electronic bands, and determine the basic characteristics of both electron and hole polarons (ground-state energy, localization lengths and effective masses) for a set of 20 materials.
We assess the limitations of the weak-coupling approach in determining the polaron effective masses: in lowest order perturbation theory, several of the studied oxide compounds go beyond the mass breakdown limit. In generalizing the isotropic electronic energy dispersion, the mass breakdown limit is reached earlier depending on the degree of electronic anisotropy present in the system. We also perform strong-coupling calculations to determine the polaron's ground-state energy and localization length. We note that for the studied oxides we reach the continuum limit inherent to the \fro model. We compare the perturbative treatment of the \fro model for anisotropic electronic dispersions with the Feynman variational path integral approach, and find them to be in reasonable agreement in the weak coupling regime. The Feynman path integral approach signals clearly where the perturbative approach fails to properly describe the polaronic behaviour.

The main results of this work are polaron properties for a set of 20 materials, obtained from the generalized \fro model, with parameters determined from first principles. 
The increased range of applicability
of the gFr model with respect to the
standard \fro model opens the possibility to analyse a wide spectrum of polarons, including
the realistic treatment of effective mass anisotropy and possible band degeneracy.

\begin{acknowledgments}
This work has been supported by the Fonds de la Recherche Scientifique (FRS-FNRS Belgium) through
the PdR Grant No. T.0103.19 - ALPS, the Fonds
de Recherche du Qu\'ebec Nature et Technologie (FRQ-NT), the Natural Sciences and
Engineering Research Council of Canada (NSERC) under grants RGPIN-2016-06666.
Computational resources have been provided by the supercomputing facilities of the
Universit\'e catholique de Louvain (CISM/UCL), the Consortium des Equipements de
Calcul Intensif en F\'ed\'eration Wallonie Bruxelles (CECI) funded by the FRS-FNRS under
Grant No. 2.5020.11 as well as the Canadian Foundation for Innovation, the Minist\'ere de
l’\'education des Loisirs et du Sport (Qu\'eec), Calcul Qu\'ebec, and Compute Canada. 
J.M.F. is supported by a Royal Society University Research Fellowship
(URF-R1-191292). 
\end{acknowledgments}


\bibliography{main} 

\end{document}


\title{
Supplementary Information: \fro polaron effective mass and localization length in cubic
materials: degenerate and anisotropic electronic bands}

\author{Bogdan Guster}
\affiliation{European Theoretical Spectroscopy Facility, Institute of Condensed Matter and Nanosciences, Universit\'{e} catholique de Louvain, Chemin des \'{e}toiles 8, bte L07.03.01, B-1348 Louvain-la-Neuve, Belgium}
\author{Pedro Melo}
\affiliation{NanoMat/Q-Mat/CESAM and European Theoretical Spectroscopy Facility, Universit\'e de Li\`ege (B5), B-4000 Li\`ege, Belgium}
\author{Bradley A. A. Martin}
\affiliation{Department of Physics, Imperial College London, Exhibition Road, London  SW7 2AZ, UK}
\author{V\'eronique Brousseau-Couture}
\affiliation{
D\'epartement de Physique et Regroupement Qu\'eb\'ecois sur les Mat\'eriaux de Pointe, Universit\'e de Montreal, C.P. 6128, Succursale Centre-Ville, Montreal, Canada H3C 3J7
}
\author{Anna Miglio}
\affiliation{European Theoretical Spectroscopy Facility, Institute of Condensed Matter and Nanosciences, Universit\'{e} catholique de Louvain, Chemin des \'{e}toiles 8, bte L07.03.01, B-1348 Louvain-la-Neuve, Belgium}
\author{Matteo Giantomassi}
\affiliation{European Theoretical Spectroscopy Facility, Institute of Condensed Matter and Nanosciences, Universit\'{e} catholique de Louvain, Chemin des \'{e}toiles 8, bte L07.03.01, B-1348 Louvain-la-Neuve, Belgium}
\author{Michel C\^ot\'e}
\affiliation{
D\'epartement de Physique et Regroupement Qu\'eb\'ecois sur les Mat\'eriaux de Pointe, Universit\'e de Montreal, C.P. 6128, Succursale Centre-Ville, Montreal, Canada H3C 3J7
}
\author{Jarvist M. Frost}
\affiliation{Department of Physics, Imperial College London, Exhibition Road, London  SW7 2AZ, UK}
\author{Matthieu J. Verstraete}
\affiliation{NanoMat/Q-Mat/CESAM and European Theoretical Spectroscopy Facility, Universit\'e de Li\`ege (B5), B-4000 Li\`ege, Belgium}
\author{Xavier Gonze}
\affiliation{European Theoretical Spectroscopy Facility, Institute of Condensed Matter and Nanosciences, Universit\'{e} catholique de Louvain, Chemin des \'{e}toiles 8, bte L07.03.01, B-1348 Louvain-la-Neuve, Belgium}\affiliation{Skolkovo Institute of
Science and Technology, Skolkovo Innovation Center, Nobel St. 3, Moscow, 143026, Russia.}

\date{\today}

\pacs{71.20.Ps, 78.20.-e, 42.70.-a}

\maketitle

\onecolumngrid

\begin{enumerate}[(I)]
    \item Supplementary theory
    \item Supplementary figures
    \item Supplementary tables
\end{enumerate}

\section{Supplementary theory}

\subsection{Polaron localization length asymptotic behavior in the uniaxial limit}


Based on the electronic effective masses, we provide analytical formulas for the asymptotic behaviour of the localization lengths in the non-degenerate anisotropic case. We determine the spherical average asymptotic behaviour in the low limit of $\mu^{*}$ (as defined in the main text), Eq. \ref{eq:ap_sph_av_l}, respectively high limit - Eq. \ref{eq:ap_sph_av_h}. Similarly, we follow with the two meaningful localization lengths $a_{p_z}$ (Eq. \ref{eq:ap_sph_av_l_z} and Eq. \ref{eq:ap_sph_av_h_z}), respectively $a_{p_{\perp}}$ (Eq. \ref{eq:ap_sph_av_l_p} and Eq. \ref{eq:ap_sph_av_h_p}), and polaron formation energy (Eq. \ref{eq:e_uni_l} and Eq. \ref{eq:e_uni_h}).

\begin{equation}
    \lim_{\mu^{*} \to 0} a_p = 2 \Big( \frac{\pi}{2} \Big)^{-5/2}m_{\perp}^{-1}\epsilon^{*} \Big( \frac{\pi\mu^{*}}{4} \Big)^{1/9}
    \label{eq:ap_sph_av_l}
\end{equation}

\begin{equation}
    \lim_{\mu^{*} \to 0} a_{p_z} = 2 \Big( \frac{\pi}{2} \Big)^{-5/2}m_{\perp}^{-1}\epsilon^{*} \Big( \frac{\pi\mu^{*}}{4} \Big)^{1/3}
    \label{eq:ap_sph_av_l_z}
\end{equation}

\begin{equation}
    \lim_{\mu^{*} \to 0} a_{p_{\perp}} = 2 \Big( \frac{\pi}{2} \Big)^{-5/2}m_{\perp}^{-1}\epsilon^{*} 
    \label{eq:ap_sph_av_l_p}
\end{equation}

\begin{equation}
    \lim_{\mu^{*} \to 0} E_{p}(\mu^{*}) = -\frac{\pi}{16}m^{*}_{\perp} (\epsilon^{*})^{-2}
    \label{eq:e_uni_l}
\end{equation}

\begin{equation}
    \lim_{\mu^{*} \to \infty} a_p = 4 \sqrt{2\pi} m^{-1}_{z} (\mu^{*}){2/3}\epsilon^{*} (\log(4\mu^{*}))^{2} (\log(4\mu^{*}))^{-2}( \log(4\mu^{2}) -2 -\log(\log(4\mu^{*}) -2 ) )^{-2/3}
    \label{eq:ap_sph_av_h}
\end{equation}

\begin{equation}
    \lim_{\mu^{*} \to \infty} a_{p_z} = 4 \sqrt{2\pi} m^{-1}_{z} \mu \epsilon^{*} (\log(4\mu^{*}))^{2} (\log(4\mu^{*}))^{-2}( \log(4\mu^{2}) -2 -\log(\log(4\mu^{*}) -2 ) )^{-1}
    \label{eq:ap_sph_av_h_z}
\end{equation}

\begin{equation}
    \lim_{\mu^{*} \to \infty} a_{p_{\perp}} = 4 \sqrt{2\pi} m^{-1}_{z} (\mu^{*})^{1/2} \epsilon^{*} (\log(4\mu^{*}))^{2} (\log(4\mu^{*}))^{-2}( \log(4\mu^{2}) -2 -\log(\log(4\mu^{*}) -2 ) )^{-1/2}
    \label{eq:ap_sph_av_h_p}
\end{equation}

\begin{equation}
    \lim_{\mu^{*} \to \infty} E_{p}(\mu^{*}) = -\frac{1}{8\pi}m_{z} (\epsilon^{*})^{-2} (\log(4) + \log(\mu^{*}))^{2}
    \label{eq:e_uni_h}
\end{equation}

In our representation inside the main text we are working at a constant density of states, thus we can define the effective mass in the following way, considering that $m_{x} = m_{y} = m_{\perp}$:

\begin{equation}
    m^{*}_{DoS} = (m_{x}m_{y}m_{z})^{1/3}
\end{equation}

\begin{equation}
    m_{z} = m^{*}_{DoS} (\mu^{*})^{-2/3}
\end{equation}

\begin{equation}
    m_{\perp} = m^{*}_{DoS} (\mu^{*})^{1/3}
\end{equation}

Consequently, the asymptotic behaviour for the different localization lengths and formation energy are defined as:

\begin{equation}
    \lim_{\mu^{*} \to 1} a_{p} = \frac{3}{2} \sqrt{2\pi} m^{-1}_{DoS} \epsilon^{*} \Big( 1 + \frac{3}{14}(\mu^{*} - 1) \Big)
\end{equation}

\begin{equation}
    \lim_{\mu^{*} \to 1} a_{p_z} = \frac{3}{2} \sqrt{2\pi} m^{-1}_{DoS} \epsilon^{*} \Big( 1 + \frac{2}{21}(\mu^{*} - 1) \Big)
\end{equation}

\begin{equation}
    \lim_{\mu^{*} \to 1} a_{p_{\perp}} = \frac{3}{2} \sqrt{2\pi} m^{-1}_{DoS} \epsilon^{*} \Big( 1 + \frac{19}{42}(\mu^{*} - 1) \Big)
\end{equation}

\begin{equation}
    \lim_{\mu^{*} \to 1} E_{p}(\mu^{*}) = -\frac{1}{6\pi}m^{*}_{DoS} (\epsilon^{*})^{-2}
\end{equation}

\subsection{Polaron properties in the ellipsoidal limit}

In the case of the tri-axial ellipsoidal limit we provide an approximate solution to the polaron localization length and formation energy. We base our estimation on the formalism provided in the asymptotic non-degenerate case where we have the exact solution for the cases of $m^{*}_{x}=m^{*}_{y} \neq m^{*}_z$ (and their commutative counterparts). Thus we estimate the average localization length in the cases where $m^{*}_{x} \neq m^{*}_{y} \neq m^{*}_z$ considering a cosine interpolation function, as defined in Eq. \ref{eq:logbeta}, while working under the constant volume condition, as defined in Eq. \ref{eq:betav}.

\begin{equation}
    a_{V} = (a_{Px}a_{Py}a_{Pz})^{1/3}
    \label{eq:betav}
\end{equation}

\begin{equation}
    \label{eqn:deltasigma}
    \delta_{\sigma} = \exp{(\sigma\sqrt{6}|c|)} - 1 ; \sigma = \pm
\end{equation}

\begin{equation}
    \ln<a> = \ln a_{V} - \frac{1}{2}\sum_{\sigma=\pm} (1 + \sigma \cos(3\theta)) (1 + \delta_{\sigma})^{1/6} S(\delta_{\sigma})
    \label{eq:logbeta}
\end{equation}

\begin{equation}
    \label{eqn:modc}
     |c| =  \Bigg( \frac{2}{3} \Big(  \ln^{2}a_{Px} + \ln^{2}a_{Py} + \ln^{2}a_{Pz} - 
                                  \ln a_{Px}\ln a_{Py}-\ln a_{Py}\ln a_{Pz}-\ln a_{Pz}\ln a_{Px} \Big) \Bigg)^{1/2}
\end{equation}

\begin{equation}
\label{eqn:dlogbeta}
\frac{ \partial \ln< a > }{ \partial a_{Pz} } =   3a_{Pz}  + 
                                                      + \frac{1}{2} \sum_{ \sigma = \pm} \Bigg\{  \frac{ \partial }{\partial a_{Pz} }( \cos(3\theta) ) \sigma ( 1 + \delta_{\sigma}  )^{1/6} S(\delta_{\sigma} ) + 
                                                      + ( 1 + \sigma \cos(3\theta) ) \frac{ \partial }{ \partial a_{Pz} } \Big[ (1 + \sigma_{\delta})^{1/6} S(\delta_{\sigma})   \Big]  \Bigg\}
\end{equation}

\begin{equation}
    \label{eqn:cos3theta}
    cos(3\theta) = \frac{4}{ (\sqrt{6}|c|)^3 } \ln \Big( \frac{a_{Px}a_{Py}}{a_{Pz}^{2}}\Big) \ln \Big( \frac{a_{Pz}a_{Px}}{a_{Py}^{2}}\Big)  \ln \Big( \frac{a_{Py}a_{Pz}}{a_{Px}^{2}} \Big)
\end{equation}

\begin{equation}
    \label{eqn:dcosthetadbz}
    \frac{ \partial }{ \partial a_{Pz} }( \cos(3\theta) ) = \cos(3\theta)\Bigg[ -\frac{3}{|c|} \frac{ \partial|c| }{ \partial a_{Pz} } 
    + a_{Pz}
    \Bigg( \ln \Big( \frac{ a_{Px}a_{Pz} }{ a_{Py}^{2} } \Big)^{-1}   + \ln \Big( \frac{ a_{Py}a_{Pz} }{ a_{Px}^{2} } \Big)^{-1} - 2\ln \Big( \frac{ a_{Px}a_{Py} }{ a_{Pz}^{2}} \Big) ^{-1}   \Bigg) \Bigg]
\end{equation}

\begin{equation}
    \label{eqn:dmocdbz}
    \frac{ \partial |c| }{ \partial a_{Pz} } = \frac{a_{Pz}}{|c|}\ln \Big( \frac{a_{Pz}^{2}}{a_{Px} a_{Py} } \Big)
\end{equation}

\begin{equation}
    \label{eqn:dds}
        \frac{ \partial }{ \partial a_{Pz} }  \Big[ ( 1 + \delta_{\sigma} )^{1/6} S( \delta_{\sigma} ) \Big] = 
         \frac{ \partial \delta_{\sigma} }{ \partial a_{Pz} } 
           \Big[ \frac{1}{6} ( 1+\delta_{\sigma} )^{-5/6}S(\delta_{\sigma}) + (1+\delta_{\sigma})^{1/6} \frac{dS}{d\delta}\Big|_{\delta_{\sigma}}  \Big]    
\end{equation}

\begin{equation}
    \label{eqn:ddeltadbz}
    \frac{ \partial \delta_{\sigma}}{ \partial a_{Pz} } = \sigma \sqrt{6} ( \delta_{\sigma} + 1 ) \frac{ \partial |c| }{ \partial a_{Pz}}
\end{equation}

\newpage

\section{Supplementary figures}

\begin{figure}[!hptb]
    \centering
    \includegraphics[width=0.65\textwidth]{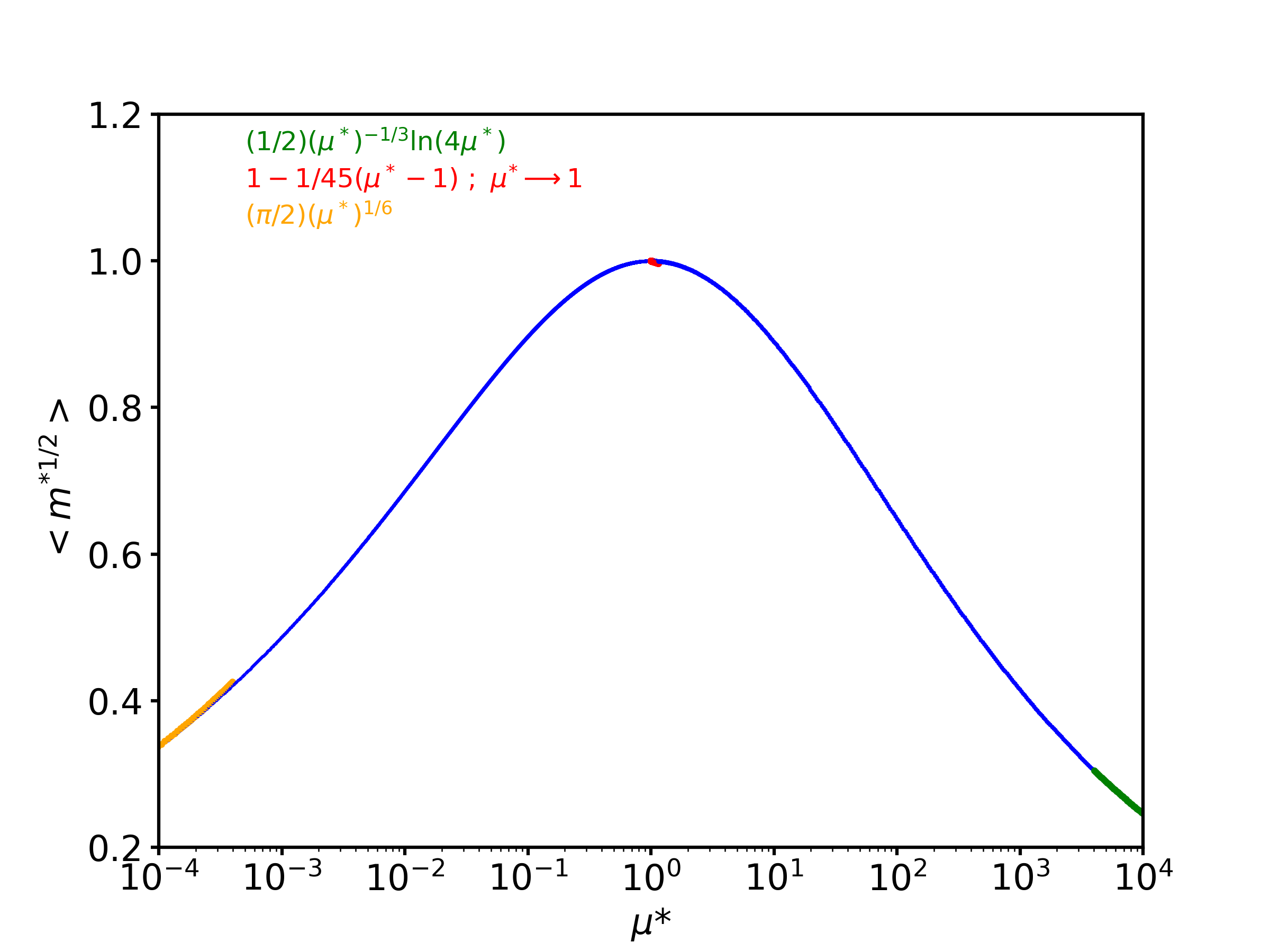}
    \caption{Average squared effective mass in the uniaxial limit calculated at constant density of states expressed in terms of $\mu^{*}$. In the low limit of $\mu^{*}$ , the asymptotic analytical behaviour is represented in orange. In the $\mu^{*}\rightarrow 1$ limit, the analytical expression is represented in red. Finally, in the large $\mu^{*}$ limit, the asymptotic behaviour analytical limit is represented in green. We consider here $\epsilon^{*} = 1$.}
    \label{fig:mave}
\end{figure}

\newpage

\section{Supplementary tables}

\begin{table}[!hptb]
\caption{$ab~initio$ DFT convergence parameters for the studied materials such as: kinetic energy cut-off in the plane-wave basis set, electronic states wavevector (k-point) sampling. }
\begin{tabular}{ | c | c | c |}
\hline\hline
Material & Ecut(Ha) & k-point sampling  \\ \hline\hline
AlAs     & 40       & 6x6x6 (x4 shifts) \\ \hline
AlN      & 35       & 6x6x6             \\ \hline
AlP      & 25       & 8x8x8 (x4 shifts) \\ \hline
AlSb     & 40       & 6x6x6 (x4 shifts) \\ \hline
BAs      & 40       & 6x6x6 (x4 shifts) \\ \hline
BN       & 35       & 8x8x8             \\ \hline
CdS      & 45       & 6x6x6 (x4 shifts) \\ \hline
CdSe     & 50       & 6x6x6 (x4 shifts) \\ \hline
CdTe     & 50       & 6x6x6 (x4 shifts) \\ \hline
GaAs     & 40       & 6x6x6 (x4 shifts) \\ \hline
GaN      & 40       & 6x6x6 (x4 shifts) \\ \hline
GaP      & 40       & 6x6x6 (x4 shifts) \\ \hline
SiC      & 35       & 6x6x6 (x4 shifts) \\ \hline
ZnS      & 40       & 6x6x6 (x4 shifts) \\ \hline
ZnSe     & 40       & 6x6x6 (x4 shifts) \\ \hline
ZnTe     & 40       & 6x6x6 (x4 shifts) \\ \hline
BaO      & 40       & 8x8x8             \\ \hline
CaO      & 40       & 8x8x8             \\ \hline
Li2O     & 50       & 8x8x8             \\ \hline
MgO      & 50       & 8x8x8 (x4shifts)  \\ \hline
SrO      & 40       & 8x8x8             \\ \hline\hline
\end{tabular}
\end{table}

\begin{table*}[!hptb]
\caption{Density functional theory calculated band parameters for triply degenerate valence bands given as Luttinger parameters, and non-degenerate anisotropic conduction bands given as electronic effective masses. In all calculation the theoretical lattice parameters were used, except for GaP where the experimental lattice parameter extrapolated at 0 K was used. The position of the band gap is also indicated.}
\begin{tabular}{|c|r|r|r|r|r|r|c|}
\hline
\hline
\multicolumn{1}{|l|}{\multirow{2}{*}{Material}} & \multicolumn{1}{l|}{\multirow{2}{*}{cell(Bohr)}} & \multicolumn{3}{c|}{VB - Luttinger Param}                           & \multicolumn{2}{c|}{CB - Eff. masses}         & \multirow{2}{*}{Gap} \\ \cline{3-7}
\multicolumn{1}{|l|}{}                          & \multicolumn{1}{l|}{}                            & \multicolumn{1}{c|}{A} & \multicolumn{1}{c|}{B} & \multicolumn{1}{c|}{C} & \multicolumn{1}{c|}{\phantom{aa}m$_{\perp}$\phantom{aa}} & \multicolumn{1}{c|}{m$_{z}$} &                      \\ 
\hline\hline
AlAs     & 10.825                          & -4.681              & -1.019              & -5.498              & 0.243                     & 0.897                     & $\Gamma$-X        \\ 
AlN      & 8.130                           & -1.435              & -0.351              & -1.793              & 0.321                     & 0.520                     & $\Gamma$-X        \\ 
AlP      & 10.406                          & -2.598              & -0.894              & -3.335              & 0.252                     & 0.809                     & $\Gamma$-X        \\ 
AlSb     & 11.762                          & -6.473              & -1.372              & -7.520              & 0.222                     & 1.142                     & $\Gamma$-L        \\ 
BaAs     & 9.088                           & -2.337              & -2.104              & -3.912              & 0.216                     & 1.094                     & ?                 \\ 
BN       & 6.746                           & -0.917              & -0.969              & -1.635              & 0.299                     & 0.895                     & $\Gamma$-X        \\
\hline
\hline
\end{tabular}
\label{tab:dft_masses}
\end{table*}

\begin{table}[!hptb]
\caption{Calculated valence band ZPR for selected materials as in Miglio et al.\cite{Miglio2020} in the generalized Fr\"ohlich formalism ($ZPR_{v}^{gFr}$) based on first principles GGA-PBE parameters. The materials considered contain triply degenerated valence bands coupled to one LO phonon branch, except BaO which consist of only one electronic band. The lattice types abbreviations correspond to zinc-blende(zb), respectively rocksalt(rs) thus falling in the cubic space group symmetry. }
\begin{tabular}{|c|c|r|r|r|r|}
\hline
\hline
Material & Lattice & \multicolumn{1}{c|}{$\omega_{LO}$ (meV)} & \multicolumn{1}{c|}{$\varepsilon_\infty$} & \multicolumn{1}{c|}{$\varepsilon_0$} & \multicolumn{1}{c|}{ZPR$_{v}^{gFr}$(meV)}      \\ 
\hline\hline
AlAs     & zb      & 47.3                           & 9.49                                & 11.51                           & 11.50                                            \\
AlP      & zb      & 59.9                           & 8.12                                & 10.32                           & 20.56                                            \\
AlSb     & zb      & 39.8                           & 12.02                               & 13.35                           & 4.00                                             \\
BAs      & zb      & 84.4                           & 9.81                                & 9.89                            & 0.53                                             \\
BN       & zb      & 161.0                          & 4.52                                & 6.69                            & 94.47                                            \\
CdS      & zb      & 34.4                           & 6.21                                & 10.24                           & 38.73                                            \\
CdSe     & zb      & 23.6                           & 7.83                                & 11.78                           & 19.69                                            \\
\hline\hline
\end{tabular}
\label{tab:dfpt_params}
\end{table}

\begin{table}[!hptb]
\caption{Convergence sampling used in the calculation of the self-energy term in the degenerate band scenario for 19 of the studied materials. $\partial^{2}\Sigma/\partial k^{2}$ was converged until a $1E-5$ Ha precision was reached.}
\begin{tabular}{|c|r|r|r|r|r|r|r|r|r|}
\hline\hline
\multirow{2}{*}{Material} & \multicolumn{3}{c|}{100}                                                       & \multicolumn{3}{c|}{110}                                                       & \multicolumn{3}{c|}{111}                                                       \\ \cline{2-10} 
                  & \multicolumn{1}{c|}{q} & \multicolumn{1}{c|}{$\theta$} & \multicolumn{1}{c|}{$\phi$} & \multicolumn{1}{c|}{q} & \multicolumn{1}{c|}{$\theta$} & \multicolumn{1}{c|}{$\phi$} & \multicolumn{1}{c|}{q} & \multicolumn{1}{c|}{$\theta$} & \multicolumn{1}{c|}{$\phi$} \\ \hline
\hline
AlAs & 240                   & 240                       & 240                     & 144                   & 144                       & 144                     & 128                   & 128                       & 128                     \\
AlP  & 144                   & 144                       & 144                     & 144                   & 144                       & 144                     & 144                   & 144                       & 144                     \\
AlSb & 240                   & 240                       & 240                     & 144                   & 144                       & 144                     & 144                   & 144                       & 144                     \\
Bas  & 128                   & 128                       & 128                     & 96                    & 96                        & 96                      & 128                   & 128                       & 128                     \\
BN   & 240                   & 240                       & 240                     & 128                   & 128                       & 128                     & 128                   & 128                       & 128                     \\
\hline\hline
\end{tabular}
\label{tab:sigma_convergence}
\end{table}

\begin{table}[!hptb]
\caption{Second order derivative of the self-energy  with respect to the wave-vector along different high-symmetry directions relevant to cubic materials. The quantity is calculated via a finite difference method around $k=0$. The quantities are expressed in atomic units.}
\begin{tabular}{|c|r|r|r|r|r|r|r|r|r|}
\hline\hline
\multirow{2}{*}{Material} & \multicolumn{3}{c|}{$\partial^{2}\Sigma/\partial k^{2}$ (100)}      & \multicolumn{3}{c|}{$\partial^{2}\Sigma/\partial k^{2}$ (110)}      & \multicolumn{3}{c|}{$\partial^{2}\Sigma/\partial k^{2}$ (111)}  \\ \cline{2-10} 
                          & \multicolumn{1}{c|}{\textbf{1}} & \multicolumn{1}{c|}{\textbf{2}} & \multicolumn{1}{c|}{\textbf{3}} & \multicolumn{1}{c|}{\textbf{1}} & \multicolumn{1}{c|}{\textbf{2}} & \multicolumn{1}{c|}{\textbf{3}} & \multicolumn{1}{c|}{\textbf{1}} & \multicolumn{1}{c|}{\textbf{2}} & \multicolumn{1}{c|}{\textbf{3}} \\ \hline\hline
AlAs                      & 0.778                           & 0.142                           & 0.142                           & 0.908                           & 0.142                           & 0.012                           & 0.951                           & 0.055                           & 0.055                           \\ \hline
AlP                       & 0.574                           & 0.164                           & 0.164                           & 0.732                           & 0.164                           & 0.006                           & 0.785                           & 0.059                           & 0.059                           \\ \hline
AlSb                      & 0.442                           & 0.078                           & 0.078                           & 0.512                           & 0.078                           & 0.008                           & 0.536                           & 0.031                           & 0.031                           \\ \hline
BAs                       & 0.008                           & 0.006                           & 0.006                           & 0.014                           & 0.006                           & 0.000                           & 0.016                           & 0.002                           & 0.002                           \\
\hline\hline
\end{tabular}
\end{table}

\begin{table*}[!hptb]
\caption{Inverse of hole polaron effective masses associated with the triply degenerate valence bands along different crystallographic directions. The quantities are expressed in atomic units.}
\begin{tabular}{|c|r|r|r|r|r|r|r|r|r|}
\hline
\hline
Material & \multicolumn{3}{c|}{1/m$^{*}_{pol}$ (100)}  & \multicolumn{3}{c|}{1/m$^{*}_{pol}$ (110)}  & \multicolumn{3}{c|}{1/m$^{*}_{pol}$ (111)}  \\ \hline\hline
AlAs     & -8.583   & -1.897 & -1.897 & -10.290  & -1.897 & -0.191 & -10.859  & -0.759 & -0.759 \\ 
AlP      & -4.621   & -1.623 & -1.623 & -6.095   & -1.623 & -0.150 & -6.585   & -0.641 & -0.641 \\ 
AlSb     & -12.503  & -2.667 & -2.667 & -14.853  & -2.667 & -0.317 & -15.636  & -1.100 & -1.100 \\ 
BAs      & -4.666   & -4.202 & -4.202 & -8.339   & -4.202 & -0.529 & -9.563   & -1.753 & -1.753 \\ 
BN       & -1.551   & -1.672 & -1.672 & -2.983   & -1.672 & -0.240 & -3.461   & -0.718 & -0.718 \\ 
CdS      & -4.906   & -0.821 & -0.821 & -5.531   & -0.821 & -0.195 & -5.740   & -0.404 & -0.404 \\
CdSe     & -12.949  & -0.951 & -0.951 & -13.738  & -0.951 & -0.161 & -14.002  & -0.424 & -0.424 \\
\hline\hline
\end{tabular}
\end{table*}

\begin{table}[!hptb]
\caption{Ratio between the electronic effective mass and polaron effective mass along different directions in the studied cubic materials characteristic to the degenerate electronic bands case.}
\begin{tabular}{|c|r|r|r|r|r|r|r|r|r|}
\hline\hline
\multirow{2}{*}{Material} & \multicolumn{3}{c|}{$m_{e}/m^{*}_{pol}$(100)}   & \multicolumn{3}{c|}{$m_{e}/m^{*}_{pol}$(110)}   & \multicolumn{3}{c|}{$m_{e}/m^{*}_{pol}$(111)}    \\ \cline{2-10} 
                          & \multicolumn{1}{c|}{\textbf{1}} & \multicolumn{1}{c|}{\textbf{2}} & \multicolumn{1}{c|}{\textbf{3}} & \multicolumn{1}{c|}{\textbf{1}} & \multicolumn{1}{c|}{\textbf{2}} & \multicolumn{1}{c|}{\textbf{3}} & \multicolumn{1}{c|}{\textbf{1}} & \multicolumn{1}{c|}{\textbf{2}} & \multicolumn{1}{c|}{\textbf{3}} \\ \hline\hline
AlAs                      & 0.917                           & 0.931                           & 0.931                           & 0.919                           & 0.931                           & 0.941                           & 0.919                           & 0.932                           & 0.932                           \\ \hline
AlP                       & 0.889                           & 0.908                           & 0.908                           & 0.893                           & 0.908                           & 0.961                           & 0.894                           & 0.916                           & 0.916                           \\ \hline
AlSb                      & 0.966                           & 0.972                           & 0.972                           & 0.967                           & 0.972                           & 0.976                           & 0.967                           & 0.972                           & 0.972                           \\ \hline
BAs                       & 0.998                           & 0.999                           & 0.999                           & 0.998                           & 0.999                           & 1.000                           & 0.998                           & 0.999                           & 0.999                           \\
\hline\hline
\end{tabular}
\end{table}


\begin{table}
\setcounter{table}{7}
\renewcommand\thetable{8.a} 

\caption{Hole polaron localization length determined in the studied materials along different high-symmetry directions relevant to cubic materials pertaining to the degenerate electronic bands case. The lengths are expressed as a function of the repetition distance between equivalent atoms in the unit cell. The repetition distance is mentioned in the last column.}
\begin{tabular}{|c|r|r|r|r|r|r|r|r|r|r|}
\hline\hline
\multirow{2}{*}{Material} & \multicolumn{3}{c|}{(100)}                                                           & \multicolumn{3}{c|}{(110)}                                                           & \multicolumn{3}{c|}{(111)}                                                           & \multicolumn{1}{c|}{\multirow{2}{*}{d(Bohr)}} \\ \cline{2-10}
                          & \multicolumn{1}{c|}{a$_1$(d)} & \multicolumn{1}{c|}{a$_2$(d)} & \multicolumn{1}{c|}{a$_3$(d)} & \multicolumn{1}{c|}{a$_1$(d)} & \multicolumn{1}{c|}{a$_2$(d)} & \multicolumn{1}{c|}{a$_3$(d)} & \multicolumn{1}{c|}{a$_1$(d)} & \multicolumn{1}{c|}{a$_2$(d)} & \multicolumn{1}{c|}{a$_3$(d)} & \multicolumn{1}{c|}{}                         \\ \hline\hline
AlAs                      & 136.43                     & 78.53                      & 78.53                      & 111.72                     & 58.73                      & 26.60                      & 117.33                     & 43.99                      & 43.99                      & 7.654                                         \\ \hline
AlP                       & 65.69                      & 44.70                      & 44.70                      & 55.18                      & 33.29                      & 14.39                      & 59.34                      & 25.09                      & 25.09                      & 7.358                                         \\ \hline
AlSb                      & 383.71                     & 218.69                     & 218.69                     & 318.70                     & 166.61                     & 80.08                      & 331.38                     & 124.90                     & 124.90                     & 8.317                                         \\
\hline\hline
\end{tabular}
\label{tab:hpll}

\end{table}
 
\begin{table}
\renewcommand\thetable{8.b}
\setcounter{table}{7}
\caption{Hole polaron formation energy in the strong coupling limit along different high-symmetry directions relevant to cubic materials pertaining to the degenerate electronic bands case.}

\begin{tabular}{|c|r|r|r|}
\hline
                   & \multicolumn{3}{c|}{E$_{SC}$(meV)}                                                                               \\ \cline{2-4} 
\multirow{-2}{*}{} & \multicolumn{1}{c|}{(100)}         & \multicolumn{1}{c|}{(110)}         & \multicolumn{1}{c|}{(111)}        \\ \hline
AlAs               & -0.13517                           & -0.20604   & -0.19736                          \\ \hline
AlP                & -0.37606                           & -0.57906   & -0.54229                          \\ \hline
AlSb               & -0.01994                           & -0.02958   & -0.02875                          \\ \hline
BAs                & -0.00023                           & -0.00031   & -0.00029                          \\ \hline
BN                 & -3.90519                           & -5.22358   & -4.90517                          \\ \hline
CdS                & -2.27488                           & -3.01849                           & -3.07203  \\
\hline
\end{tabular}
\end{table}

\begin{table}[!hptb]
\caption{Electron polaron localization length expressed in units of the repetition distance between equivalent atoms. a$_{p}$(d) is the spherical average determined by Eq. 108 where the electronic effective masses expressed in Table \ref{tab:dft_masses} have been used. a$_p{_\perp}$ is the in-plane localization length.  a$_p{_z}$ is the out-of-plane localization length. The radii are determined in the \fro model.}
\begin{tabular}{|c|r|r|r|r|c|}
\hline\hline
Material & \multicolumn{1}{c|}{a$_p$(d)} & \multicolumn{1}{c|}{a$_p{_{\perp}}$(d)} & \multicolumn{1}{c|}{a$_p{_{z}}$(d)} & \multicolumn{1}{c|}{Ratio} & Shape \\ \hline\hline
AlAs & 13.81   & 16.10  & 10.15  & 0.63 & oblate    \\ \hline
AlN  & 28.70   & 30.39  & 25.59  & 0.84 & oblate    \\ \hline
AlP  & 12.04   & 13.81  & 9.14   & 0.66 & oblate    \\ \hline
AlSb & 18.92   & 22.94  & 12.87  & 0.56 & oblate    \\ \hline
BAs  & 259.24  & 313.63 & 177.12 & 0.56 & oblate    \\ \hline
BN   & 6.37    & 7.25   & 4.92   & 0.68 & oblate    \\ \hline
CdS  & 63.62   & -      & -      & -    & spherical \\ \hline
CdSe & 207.27  & -      & -      & -    & spherical \\ \hline
CdTe & 259.36  & -      & -      & -    & spherical \\ \hline
GaAs & 6439.74 & -      & -      & -    & spherical \\ \hline
GaN  & 59.65   & 59.65  & 59.65  & 1.00 & spherical \\ \hline
GaP  & 13.15   & 15.75  & 9.17   & 0.58 & oblate    \\ \hline
SiC  & 10.67   & 12.14  & 8.26   & 0.68 & oblate    \\ \hline
ZnS  & 50.60   & -      & -      & -    & spherical \\ \hline
ZnSe & 128.23  & -      & -      & -    & spherical \\ \hline
ZnTe & 219.80  & -      & -      & -    & spherical \\ \hline
BaO  & 1.09    & 1.24   & 0.83   & 0.67 & oblate    \\ \hline
CaO  & 1.04    & 1.19   & 0.79   & 0.66 & oblate    \\ \hline
Li2O & 2.37    & 2.57   & 2.03   & 0.79 & oblate    \\ \hline
MgO  & 7.26    & -      & -      & -    & spherical \\ \hline
SrO  & 0.98    & 1.11   & 0.75   & 0.68 & oblate    \\ \hline\hline
\end{tabular}
\label{tab:epll}
\end{table}

\begin{table}[!hptb]
\caption{Electron polaron localization length expressed in units of the repetition distance between equivalent atoms. a$_{p}$(d) is the spherical average determined by Eq. 37 where the electronic effective masses expressed in Table \ref{tab:dft_masses} have been used. a$_p{_\perp}$ is the in-plane localization length.  a$_p{_z}$ is the out-of-plane localization length. The radii are determined in the Feynman model.}
\begin{tabular}{|c|r|r|r|r|c|}
\hline\hline
Material & \multicolumn{1}{c|}{a$_p$(d)} & \multicolumn{1}{c|}{a$_p{_{\perp}}$(d)} & \multicolumn{1}{c|}{a$_p{_{z}}$(d)} & \multicolumn{1}{c|}{Ratio} & Shape     \\ \hline\hline
AlAs     & 16.72                  & 28.72                   & 5.66                    & 0.20                      & oblate    \\ \hline
AlP      & 13.96                  & 22.64                   & 5.31                    & 0.23                      & oblate    \\ \hline
AlSb     & 23.26                  & 45.97                   & 5.96                    & 0.13                      & oblate    \\ \hline
BAs      & 81.55                  & 160.16                  & 21.14                   & 0.13                      & oblate    \\ \hline
BN       & 8.57                   & 13.47                   & 3.47                    & 0.26                      & oblate    \\ \hline
CdS      & 40.82                  & -                       & -                       & -                         & spherical \\ \hline
CdSe     & 146.15                 & -                       & -                       & -                         & spherical \\
\hline\hline
\end{tabular}
\label{tab:epll_feynman}
\end{table}

\begin{table}[!hptb]
\caption{Calculated conduction band ZPR for selected materials as in Miglio et al.\cite{Miglio2020} in the generalized Fr\"ohlich formalism ($ZPR_{v}^{gFr}$) based on first principles GGA-PBE parameters and compared to the Feynman variational approach for both isotropic and anisotropic non-degenerate electronic bands.}
\begin{tabular}{|c|r|r|r|r|r|}
\hline\hline
\multirow{2}{*}{Material} & \multicolumn{1}{c|}{\fro} & \multicolumn{3}{c|}{Feynman}                                                  & \multicolumn{1}{c|}{\multirow{2}{*}{R.E.(\%)} } \\ \cline{2-5}
                          & \multicolumn{1}{c|}{ZPR$_c^{gFr}$(meV)}  & \multicolumn{1}{c|}{F$_{avg}$(meV)} & \multicolumn{1}{c|}{F$_{\perp}$(meV)} & \multicolumn{1}{c|}{F$_{z}$(meV)} & \multicolumn{1}{c|}{}                   \\ \hline\hline
AlAs                      & -8.8                          & -9.6                     & -7.3                      & -14.1                  & -9.4\%                                         \\ \hline
AlP                       & -14.0                         & -15.1                    & -11.9                     & -21.4                  & -7.6\%                                         \\ \hline
AlSb                      & -3.6                          & -4.1                     & -2.9                      & -6.5                   & -15.0\%                                        \\ \hline
BAs                       & -0.5                          & -0.6                     & -0.4                      & -0.9                   & -14.2\%                                        \\ \hline
BN                        & -67.9                         & -72.6                    & -58.3                     & -101.3                 & -7.0\%                                         \\ \hline
CdS                       & -14.9                         & -15.0                    & -15.0                     & -15.0                  & -0.5\%                                         \\
\hline\hline
\end{tabular}
\end{table}

\begin{table}[!hptb]
\caption{Calculated conduction band polaron effective masses for selected materials as in Miglio et al.\cite{Miglio2020} in the generalized cubic Fr\"ohlich formalism based on first principles GGA-PBE parameters and compared to the Feynman variational approach for both isotropic and anisotropic non-degenerate electronic bands.}
\begin{tabular}{|c|r|r|r|r|r|r|}
\hline\hline
\multirow{2}{*}{Material} & \multicolumn{2}{c|}{\fro}                        & \multicolumn{2}{c|}{Feynman}                          & \multicolumn{2}{c|}{R.E.}                            \\ \cline{2-7} 
                          & \multicolumn{1}{c|}{m$^{*}_{pol_{\perp}}$[a.u.]} & \multicolumn{1}{c|}{m$^{*}_{pol_{z}}$[a.u.]} & \multicolumn{1}{c|}{m$^{*}_{pol_{\perp}}$[a.u.]} & \multicolumn{1}{c|}{m$^{*}_{pol_{z}}$[a.u.]} & \multicolumn{1}{c|}{ /m$^{*}_{pol_{\perp}}$[a.u.] }  & \multicolumn{1}{c|}{m$^{*}_{pol_{z}}$[a.u.]} \\ \hline\hline
AlAs                      & 0.25181                    & 0.91550                 & 0.24901                    & 0.94087                 & 1.1\%                      & -2.8\%                  \\ \hline
AlP                       & 0.26372                    & 0.83222                 & 0.26024                    & 0.85736                 & 1.3\%                      & -3.0\%                  \\ \hline
AlSb                      & 0.22561                    & 1.15212                 & 0.22422                    & 1.17299                 & 0.6\%                      & -1.8\%                  \\ \hline
BAs                       & 0.21668                    & 1.09474                 & 0.21660                    & 1.09608                 & 0.0\%                      & -0.1\%                  \\ \hline
BN                        & 0.32467                    & 0.94301                 & 0.31661                    & 0.98838                 & 2.5\%                      & -4.8\%                  \\ \hline
CdS                       & 0.12688                    & 0.12688                 & 0.12622                    & 0.12622                 & 0.5\%                      & 0.5\%                   \\
\hline\hline
\end{tabular}
\label{tab:epem}
\end{table}

\begin{table}[!hptb]
\caption{Feynman path integral variational method parameters for the non-degenerate anisotropic case.}
\begin{tabular}{|c|r|r|r|r|}
\hline\hline
Materials & \multicolumn{1}{c|}{v$_{\perp}$} & \multicolumn{1}{c|}{w$_{\perp}$} & \multicolumn{1}{c|}{v$_{z}$} & \multicolumn{1}{c|}{w$_{z}$} \\ \hline\hline
AlAs      & 3.01564                 & 2.98092                 & 3.03038                 & 2.96301                 \\ \hline
AlP       & 3.02014                 & 2.97542                 & 3.03662                 & 2.95554                 \\ \hline
AlSb      & 3.00726                 & 2.99193                 & 3.01659                 & 2.97979                 \\ \hline
BAs       & 3.00054                 & 2.99946                 & 3.00569                 & 3.00326                 \\ \hline
BN        & 3.03711                 & 2.95497                 & 3.06588                 & 2.92091                 \\ \hline
CdS       & 3.04481                 & 2.94577                 & 3.04481                 & 2.94576                 \\ \hline
CdSe      & 3.02365                 & 2.97114                 & 3.02366                 & 2.97114                 \\
\hline\hline
\end{tabular}
\end{table}

\bibliography{main}